\documentclass[article]{aa}

\usepackage{longtable,lscape}

\usepackage{natbib}
\bibpunct{(}{)}{;}{a}{}{,}
\usepackage{graphicx}

\usepackage{graphicx}
\usepackage{txfonts}
\usepackage{natbib}
\usepackage[colorlinks=true,citecolor=blue]{hyperref}
\usepackage{color}
\usepackage{xspace}
\usepackage{dcolumn}
\usepackage{longtable}
\usepackage{lscape}
\usepackage{soul}        

\usepackage{rotating}
\usepackage{afterpage}

\newcommand{\MJup}{M$_{\mathrm{Jup}}$\xspace}

\newcommand{\mic}{$\mu$m\xspace}
\newcommand{\as}{\hbox{$^{\prime\prime}$}\xspace}
\def\tablebib#1{\par\vspace*{2ex}%
 \parbox{\hsize}{\leftskip0pt\rightskip0pt
 {\noindent\small\textbf{References.}~#1\par}}}

\usepackage{txfonts}
\begin{document}
\title{Astrophysical false positive{s} in direct imaging for exoplanets: 
a white dwarf close to a rejuvenated star
\thanks{Based on observations collected at La Silla and Paranal Observatory, 
ESO (Chile): Programs  184.C-0567 (NAOS-CONICA), 184.D-1151 (EFOSC), 60.A-9036 
(HARPS), 089.C-0665 (SINFONI). Based on the All Sky Automated Survey (ASAS) 
photometric data.}}

   \author{A. Zurlo\inst{1,2}, A. Vigan\inst{1}, J. Hagelberg\inst{3}, S. Desidera\inst{2}, G. Chauvin\inst{4}, 
 J.~M. Almenara\inst{1},\\ K. Biazzo\inst{5}, M. Bonnefoy\inst{6}, J.~C. Carson\inst{7}, E. Covino\inst{5}, P. Delorme\inst{4}, V. D'Orazi\inst{8,9},  R. Gratton\inst{2}, D. Mesa\inst{2}, S. Messina\inst{10}, C. Moutou\inst{1}, D. Segransan\inst{3}, M. Turatto\inst{2}, S. Udry\inst{3}, F. Wildi\inst{3} }

   \institute{\inst{1} Aix Marseille Universit\'e, CNRS, LAM (Laboratoire d'Astrophysique de Marseille) UMR 7326, 13388, Marseille, France\\
   \inst{2} INAF-Osservatorio Astronomico di Padova, Vicolo dell'Osservatorio 5, 35122, Padova, Italy \\               
\inst{3} Geneva Observatory, University of Geneva, Chemin des Mailettes 51, 1290, Versoix, Switzerland\\   
\inst{4} Institut de Plan\'etologie et d'Astrophysique de Grenoble, UJF, CNRS, 414 rue de la piscine, 38400, Saint Martin d'H\`eres, France\\  
 \inst{5} INAF-Osservatorio Astronomico di Capodimonte, salita Moiariello 16, 80131, Napoli, Italy\\
    \inst{6} Max-Planck-Institute fur Astronomie, Konigsthul 17, 69117, Heidelberg, Germany\\
   \inst{7} Department of Physics \& Astronomy, College of Charleston, 58 Coming St., Charleston, SC 29424, USA \\
                  \inst{8} Department of Physics \& Astronomy, Macquarie University, Balaclava Rd, North Ryde, Sydney, NSW 2109, Australia \\
\inst{9} Monash Centre for Astrophysics, Monash University, School of Mathematical Sciences, Building 28, Monash University, Melbourne, VIC 3800, Australia\\
\inst{10} INAF-Osservatorio Astrofisico di Catania, Via S. Sofia 78, 95123, Catania, Italy\\}
   \date{Accepted for publication in A\&A; 14 April 2013 }


  \abstract
   {As is the case for all techniques involved in the research for exoplanets, direct imaging 
   has to take into account the probability of so-called astrophysical 
   \textit{false positives}, which are phenomena that mimic the signature of objects we are seeking. }
   {In this work we aim to present a case of {{a}} false positive found during a direct 
   imaging survey conducted {{with}} VLT/NACO. A promising exoplanet candidate was detected 
   around the {{K2-type}} star HD\,8049 in July 2010. Its contrast of $\Delta H$ = 7.05 at 1.57 
   arcsec allowed us to guess the presence of a 35~\MJup companion at 50 projected AU, for the nominal system age and heliocentric distance.}
   {To check whether it was gravitationally bound to the host star, as opposed to an unrelated background object, we re-observed the system one year later and concluded a high probability of a bound system.   We also used radial velocity measurements of the 
   host star, spanning a time range of $\sim$ 30~yr, to constrain the companion's mass and orbital properties, as well as to probe the host star's spectral age indicators and general spectral energy distribution. We also obtained {{for the companion}} $U$-band imaging with EFOSC and $NIR$ spectroscopy.}
   {Combining all these information we conclude that the companion of HD\,8049 is a white dwarf (WD), with temperature $T_{\mathrm{eff}} = 18800 \pm 2100$~K and mass $M_{\mathrm{WD}} = 0.56 \pm 0.08~M_{\odot}$.  The significant radial velocity trend coupled with the imaging data indicate that the most probable orbit has a semimajor axis of about 50 AU.  The discrepancy between the age indicators suggested against a 
   \textit{bona-fide} young star.  The moderately 
   large level of chromospheric activity and fast rotation, mimicking the properties of 
   a young star, might be induced by the exchange of mass with the progenitor of the WD. 
   This example demonstrates some of the challenges in determining accurate age estimates and identifications of faint companions. }
   {}

   \keywords{Techniques: high angular resolution - spectroscopic -
              Stars: white dwarfs - brown dwarfs - individual: HD\,8049 -  binaries: visual}

\titlerunning{Astrophysical false positive in the NACO}
\authorrunning{Zurlo et al.}
\maketitle
%

\section{Introduction}

In the search for extrasolar planets, we are frequently faced  with objects or physical 
phenomena that mimick the signature of a substellar object --- the so-called astrophysical 
\textit{false positives}. The most {commonly} used techniques in the search for exoplanets are 
the radial velocit{y}, transit, TTV (Transit Timing Variation), microlensing and direct imaging {techniques}. A large 
fraction of the planets discovered so far has been detected through the radial 
velocity technique. Rotational modulations of star spots and magnetic activity 
cycles might produce spurious, periodic, radial velocity signals that mimic planet 
signatures \citep[see e.g.][]{Que01,2011arXiv1107.5325L}. A number of controversial 
cases have been reported in the literature 
\citep[e.g.][]{2008Natur.451...38S,2008A&A...489L...9H,2010A&A...512A..45H,2010A&A...513L...8F}.
Simultaneous monitoring of activity indicators, spectral line profile changes and 
photometric variations {may allow one} to recognize the true origin of the radial velocity 
variations and {to} at least partially account for them
\citep[see e.g.][]{2011A&A...528A...4B,2010A&A...520A..53L,2012Natur.491..207D}.
   
Transit searches are even more heavily plagued by false positives.  Various kinds of eclipsing binaries can mimick
 a transiting planet photometric curve. Such false positives can often outnumber the true exoplanet detections \citep{Brown2003}. The {phenomenon} of blended eclipsing binaries is the most severe, as 
described in \citet{Odo06}.  To help mitigate these adverse effects, various observational diagnostics, based on the photometric 
light curves or on additional spectroscopic or photometric observations, have been 
developed. Statistical tools can also be exploited to explore all possible scenarios 
of binary systems and the probability of their occurrence 
\citep[e.g. BLENDER code,][PASTIS code, D\'iaz et al., in preparation]{torres2011}. 

Timing variations in eclipsing binaries might be signatures of circumbinary planets, but 
alternative explanations cannot be firmly ruled out \citep{2012arXiv1211.5356Z}.  A recent case of TTVs induced by stellar activity (star spot features modulating the brightness of the host) instead of the gravitational influence of a planet is described in \citet{Barros2013}.  Finally, a case of a possible microlensing planet detection was recently revealed to be a false 
positive, resulting from the variability of the lensed star \citep{Gould2012}.
 
Concerning the direct imaging technique, the subject of this work, it may occur that the 
detected candidate companion is not gravitationally bound to the host star. Detection 
of unbound objects is frequent since direct imaging surveys {typically have better sensitivities to} to planets 
far from the host star, {where a higher contrast can be reached} \citep[see e.g.][]{Chauvin2,vigan2012}. If the follow-up observation is 
close in time to the first epoch and the displacement of the star on the sky due to 
proper motion is small, {it may be challenging to} determine with a high probability {whether an} object 
is bound. One example is given by \citet{2009ApJS..181...62M} where an object that was originally {thought}
 to be compatible with being bound, based on the proper motion test, was determined through spectroscopy to be 
a background M-star. Moreover, the case of the false positive 
object around IM Lupi \citep{Maw12} teaches that, when the distance determination
is complex, {it is possible to miscalculate the relative certainties of a bound versus unbound interpretation.}  Secondly, even after a candidate companion is shown to be bound, the system might prove to be a stellar binary, which is the case of the object of our study.

Finally, in order to meaningfully characterize the companion, the system age must be accurately determined, as the luminosity of 
substellar objects depends {strongly} on the age of the system \citep[younger age 
corresponds to a brighter companion magnitude, see e.g.][]{Baraffe2003,Madhusudhan2011}; 
spurious young stars might therefore {lead to} poorly classified companions.   In this paper we present the case of such a spurious young star, along with an originally misidentified substellar candidate, found in the NACO 
Large Program \citep[NACO-LP\footnote{ESO program 184.C-0567, PI J.-L. Beuzit, ``Probing 
the Occurrence of Exoplanets and Brown Dwarfs at Wide Orbits''.};][]{chauvin2010}.

The object of our study, HD\,8049, was observed in July 2010. The target is a {solar neighborhood K2-star}, with a distance of 33.6~pc.
The companion found from the direct imaging observations had an $H$-band magnitude consistent with  a substellar mass, based on
the nominal age of the central star. However, 
after a deeper analysis, presented in this paper, it {is concluded} to be a white dwarf (WD). The 
case is similar to that of the object orbiting Gliese~86 \citep[see][]{Mug05}, initially 
considered a brown dwarf by \citet{Els01}. Other cases of close WD companions 
around HD\,147513 and HD\,27442 are described in \citet{portodemello1997} and 
\citet{Chauvin2007} respectively.

Sometimes, as in the case presented here, a  false positive turns out to be an interesting object: {among known nearby WD} this target is in fact one of the hottest. Because it is part of a close binary system, it {yields insights into} how those types of object interact with their companions. 

This work is structured as follows: we present the observations in 
Section~\ref{sec:observations_data_reduction}; we show the companion characteristics 
in Section~\ref{sec:companion}; we summarize the stellar properties and how they are 
influenced by the companion in Section~\ref{sec:star}; the properties of the binary 
system are discussed in Section~\ref{sec:system}; and conclusions are drawn in 
Section~\ref{sec:conclusions}.

\section{Observations and data reduction}
\label{sec:observations_data_reduction}
HD\,8049 was observed with several instruments and techniques. All data we collected for this analysis
are listed in Table \ref{t:obs} along with the related instrumentation, date of observation and technique.

\begin{table}
\caption{Resuming table of the observations of HD\,8049.}
\label{t:obs}
\begin{tabular}{lll}

\hline
\hline
Instrument      &  Date                       & Type of Data \\
\hline
CORAVEL         & June 14 UT, 1983            & 1 RV p.,{$V$ spectrum} \\
                & December 31 UT, 1983	      & 1 RV p.,{$V$ spectrum} \\
                & November 30 UT, 1985	      & 1 RV p.,{$V$ spectrum} \\
CORALIE(98)     & from December 26 UT, 2000   & 43 RV points \\
                & to January 21 UT, 2007      & \\
HARPS           & September 15 UT, 2004       & 1 RV p.,{$V$ spectrum} \\
CORALIE(07)     & from July 3 UT, 2007        & 20 RV points \\
                & to July 8 UT, 2012	      &  \\
NACO         	& July 30 UT, 2010            &{$H$-band imaging} \\
                & July 28 UT, 2011            &{$H$-band imaging} \\
EFOSC           & December 21 UT, 2011        &{$U$-band imaging},\\
                &                             &{$V$ spectrum} \\
SINFONI         & August 28 UT, 2012          &{$JHK$ spectra} \\
\hline
\end{tabular}

\end{table}

\subsection{NACO observations}
\label{sub:naco}

HD\,8049 was observed as part of the NACO Large Program (NACO-LP) imaging survey. We used 
VLT/NACO \citep{lenzen2003,rousset2003} high-contrast Adaptive Optics (AO) system with 
the $H$-band (1.33--1.99~\mic) and the S13 camera, which provides a spatial
sampling of $\sim$13~mas/pixel and a field of view (FoV) of $14\as \times 14\as$. 
The target was observed in pupil-tracking mode to allow implementation of angular 
differential imaging \citep[ADI,][]{marois2006} data analysis techniques. The 
observing sequence consisted of a series of unsaturated images of the point spread 
function (PSF), which served as reference for differential photometry and astrometry, 
and a series of deep saturated exposures optimized for the detection of faint 
companions. Data reduction for saturated and unsaturated sequences followed standard 
procedures (flat--fielding, sky subtraction, bad pixel correction). The saturated 
sequence was then analysed with two methods: (1) the LOCI algorithm \citep{lafreniere2007} 
with a separation criterion $N_{\delta} = 0.75$~FWHM, and (2) a spatial filtering of 
all images in a 5$\times$5~FWHM box followed by de-rotation and median-combination. 
Differential astrometry was performed using fitting of a 2D Gaussian profile on the 
primary star and on the point sources detected in the reduced images.

\begin{figure}[h]
\begin{center}
\includegraphics[height=5cm]{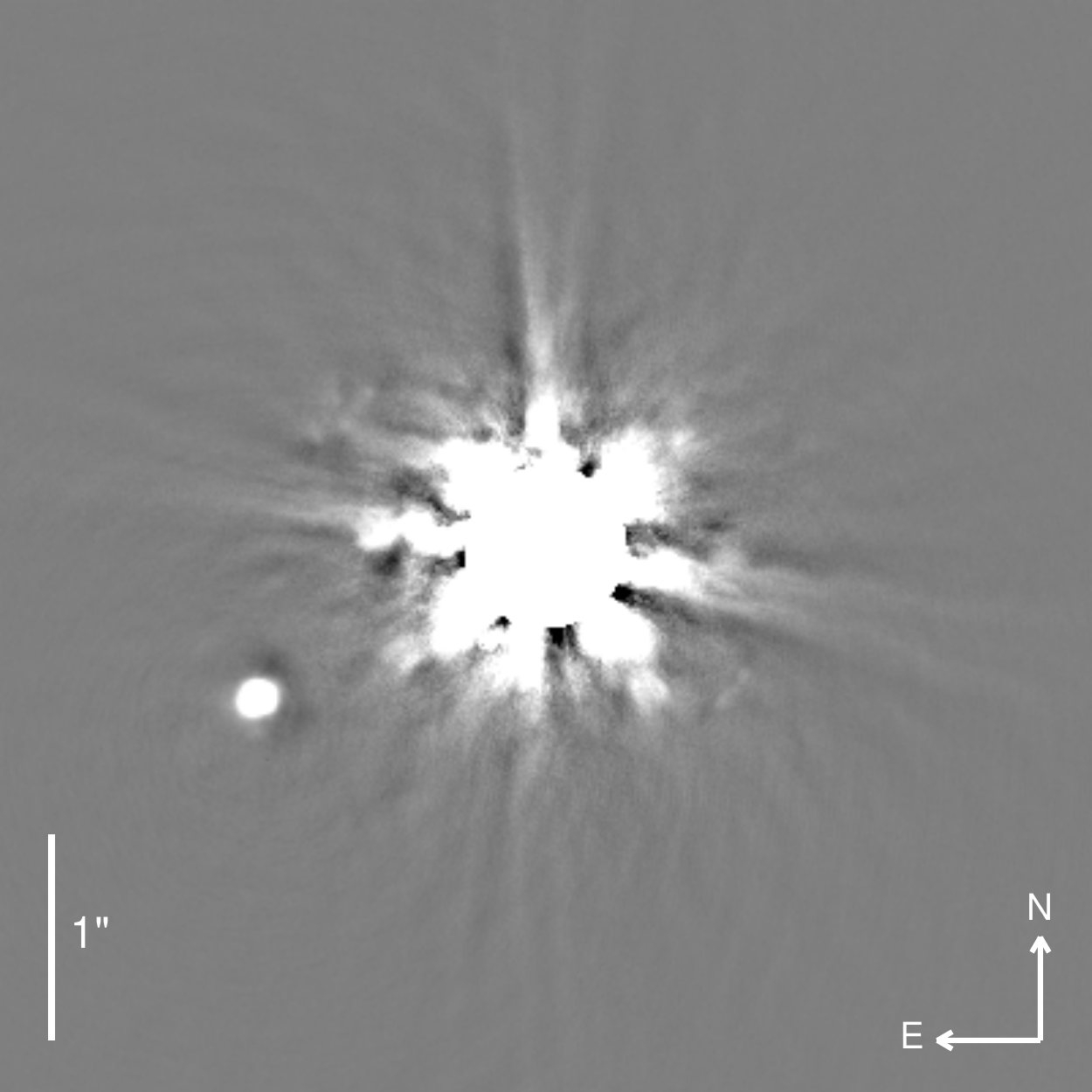}
\caption{{VLT/NACO, $H$-band image (1.65~$\mu$m) of the companion. The frame was generated from the median combination of processed images having undergone spatial filtering (5$\times$5$\lambda/D$ box) and de-rotation. The central core of the PSF is saturated and has been masked in this image.}}
\label{f:naco}
\end{center}
\end{figure}

The target was first observed on July 30 UT, 2010. {A} faint
point source (hereafter HD\,8049\,B) at a separation of
$\rho$ = $1.566~\pm~0.006~\arcsec$ and with a position angle of 
$\theta$ = $118.4~\pm~0.2$~deg was detected with a magnitude difference 
$\Delta H$ = $7.1\pm0.1$~mag (see Figure \ref{f:naco}). Follow-up observations 
on July 28 UT, 2011 revealed that the point source was co-moving with the
primary. Thanks to the high proper motion of the central star
($\mu_{\alpha} = 65.99\pm1.18$~mas/yr and $\mu_{\delta} =
-240.99\pm0.98$~mas/yr), a $\chi^2$ probability test on $\Delta\alpha$
and $\Delta\delta$ with respect to the star at two epochs rejected the
possibility (at 99\% certainty) that the object was  a background source. {The measured contrast, along with our original age estimate and the measured astrometry, led us to conclude} the presence 
of a 35~\MJup companion at 50~AU projected separation.

\subsection{CORAVEL, CORALIE and HARPS high resolution spectroscopy}
\label{sub:spectra}

HD\,8049 was observed as part of a long term extra-solar planet
search program started in June 1998 with the CORALIE high-resolution
fibre-fed echelle spectrograph mounted on the 1.2~m Euler Swiss
telescope at La Silla (ESO, Chile). In addition to the CORALIE data, we obtained a
spectrum with HARPS.  Finally, we used the CORAVEL database \citep{Baranne1996} and extracted 3 radial velocity (RV) observations with a
300~m/s precision \citep{Baranne1996}. This
enabled us to have a total time-span of nearly 29 years. 

\begin{figure}[h]
\begin{center}
\includegraphics[height=7cm]{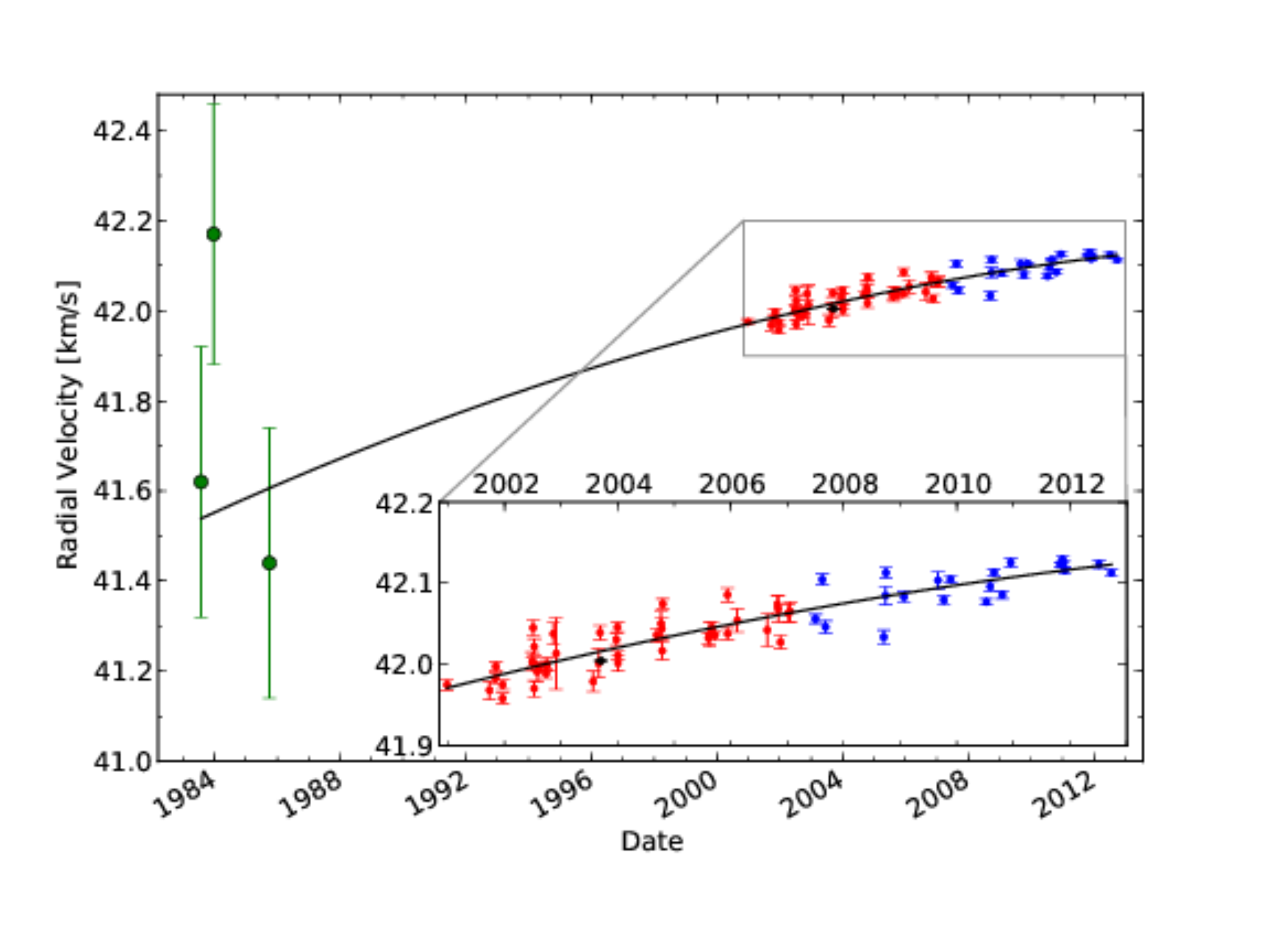}
\caption{{Radial velocity of the star measured with four instruments: CORAVEL (green), CORALIE98 (red), CORALIE07 (blue) and HARPS (black).  The solid black line is a least-squares fit of a quadratic RV trend. While the drift information alone is consistent with either a BD or WD interpretation, the inclusion of the separation information, as gathered from NACO, allows us to rule out the BD possibility.}}
\label{f:rv}
\end{center}
\end{figure}

For these
three instruments, radial velocities were computed by cross-correlating
the observed stellar spectra with a binary mask, whose non-zero zones
correspond to the theoretical positions and widths of stellar
absorption lines at zero velocity. For HARPS and CORALIE, the
instrumental velocity drifts were monitored and corrected using the
``simultaneous thorium referencing technique'' with dual fibres
\citep[more details in][]{Baranne1996}, whereas for CORAVEL a standard
radial velocity star (HD\,168454) was used for calibration. Radial
velocity points are shown in Figure \ref{f:rv}. The trend in the
radial velocity points is likely produced by the influence of a more
massive object than a brown dwarf. Additionally, activity indicators (S
index, $\log R_{\mathrm{HK}}$) and the projected rotational velocity were also
measured on the CORALIE and HARPS spectra.

\subsection{NTT/EFOSC}
\label{sub:uphoto}

\begin{figure}[h]
\begin{center}
\includegraphics[height=5cm]{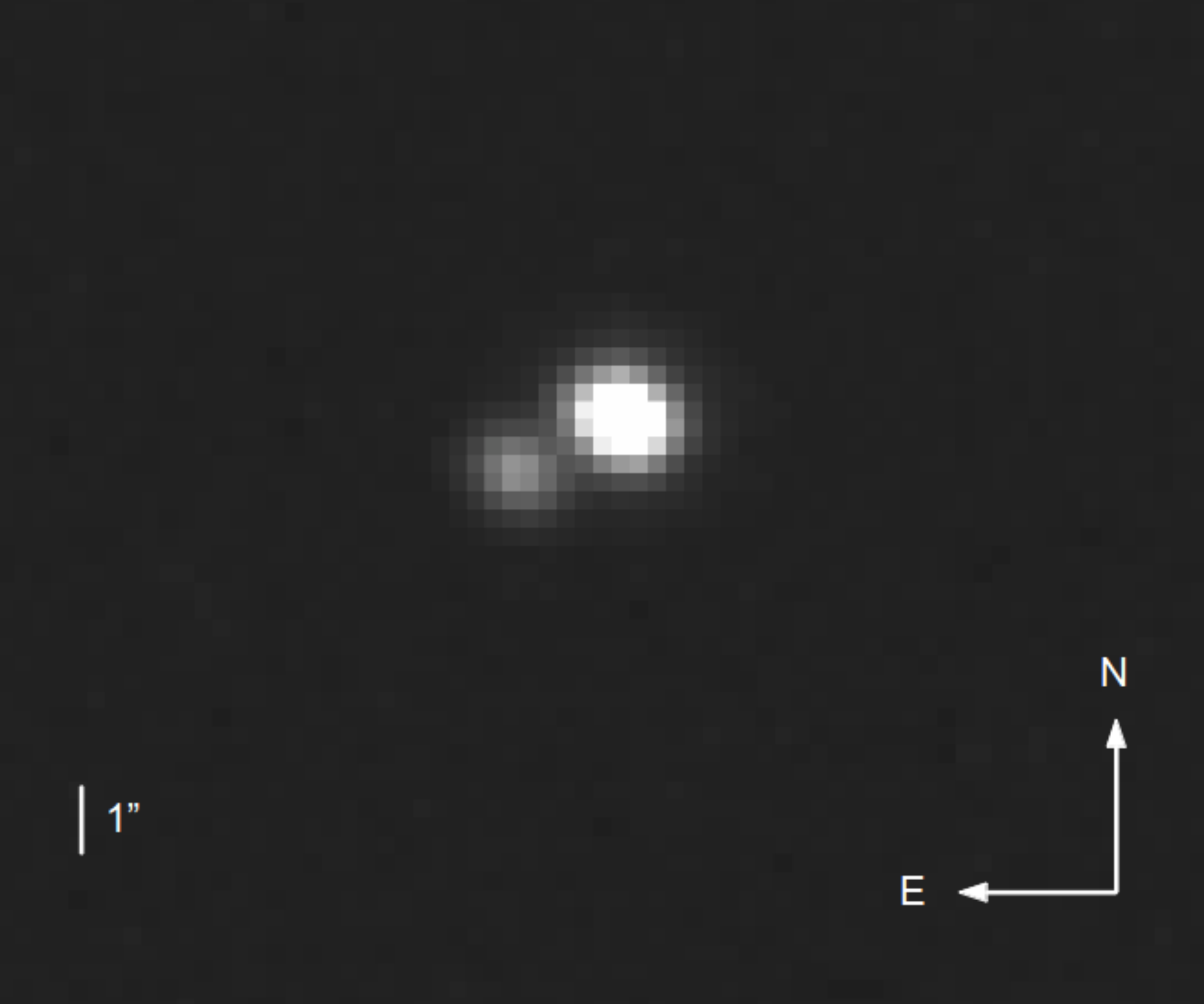}
\caption{$U$-band EFOSC image of the binary system. HD\,8049\,B, in the south
east side, is 1.7~mag fainter.}
\label{f:uband}
\end{center}
\end{figure}

We obtained 10 images of HD\,8049 on December 21 UT, 2011, with
EFOSC2 at the NTT (ESO, Chile). Each frame was taken with
the Johnson $U$ filter, a 0.2~s exposure to avoid saturation,
2$\times$2 binning and seeing of 0.8$\arcsec$. We reduced the $U$-band
images with standard IRAF\footnote{IRAF is distributed by the National Optical
Astronomy Observatory, which is operated by the Association of
Universities for Research in Astronomy (AURA) under cooperative
agreement with the National Science Foundation.} tasks. We fitted the two objects with 
Moffat profiles and performed aperture photometry on 
each star after subtracting contributed flux from the neighbor star. The 
resolved binary system had a magnitude difference of $\Delta U$ = $1.714\pm0.023$~mag, 
as shown in Figure \ref{f:uband}. The difference was calculated taking 
into account the correction for the $U-B$ color index of the two stars 
using equation (9) of the EFOSC2 manual\footnote{EFOSC2 manual, version 
3.6. ESO user facilities.}. We used for the K star the value listed in 
Table~15.7 of \citet{Allen} and for the WD we considered the 
colors in the range of effective temperature, $T_{\mathrm{eff}}$, from 15000~K to 25000~K 
(see Sec~\ref{sec:wd}). 
The measured values of the projected separation, $\rho=1.56\pm 0.02\arcsec$, and 
the position angle, $\theta=118\pm0.5\deg$, were compatible with NACO measurements.

During the same night, we performed spectroscopy of the system. The {n.11} grism and the 1$\arcsec$ slit were used, providing a resolution R $\sim 360$ 
over the range between 3500--7400~$\AA$. The slit was oriented along
the two stars (i.e. PA = 121 deg). We obtained two frames of 8~s exposure.  

The spectral reduction was done
using standard IRAF tasks. The raw data were bias subtracted and
flat-field corrected. The spectra were extracted with the IRAF task
\textit{apall}, wavelength calibrated using arc lamp exposures, and
cross-correlated with the sky lines. The relative flux calibration
was performed by deriving the sensitivity curve for the instrument set-up
from the spectroscopic standard star L745-46A \citep{Hamuy1994}.

We found that a low S/N signature of HD\,8049\,B was present in the blue 
part of the spectrum; its contribution is completely blended with 
that of the primary star.

\subsection{VLT/SINFONI}
\label{sub:sinfoni}

\begin{figure*}
\begin{center}
\includegraphics[width=0.48\textwidth]{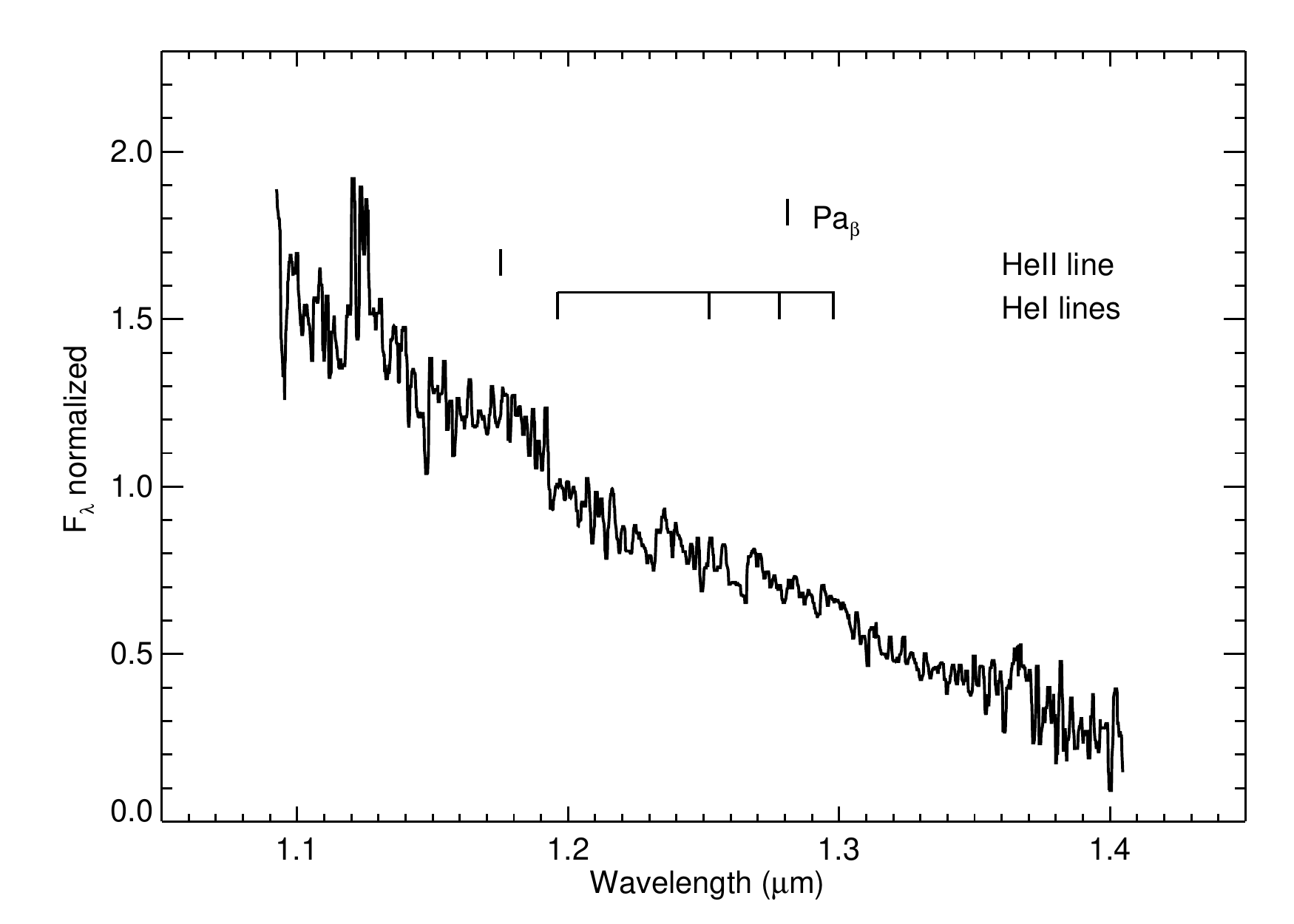}
\includegraphics[width=0.48\textwidth]{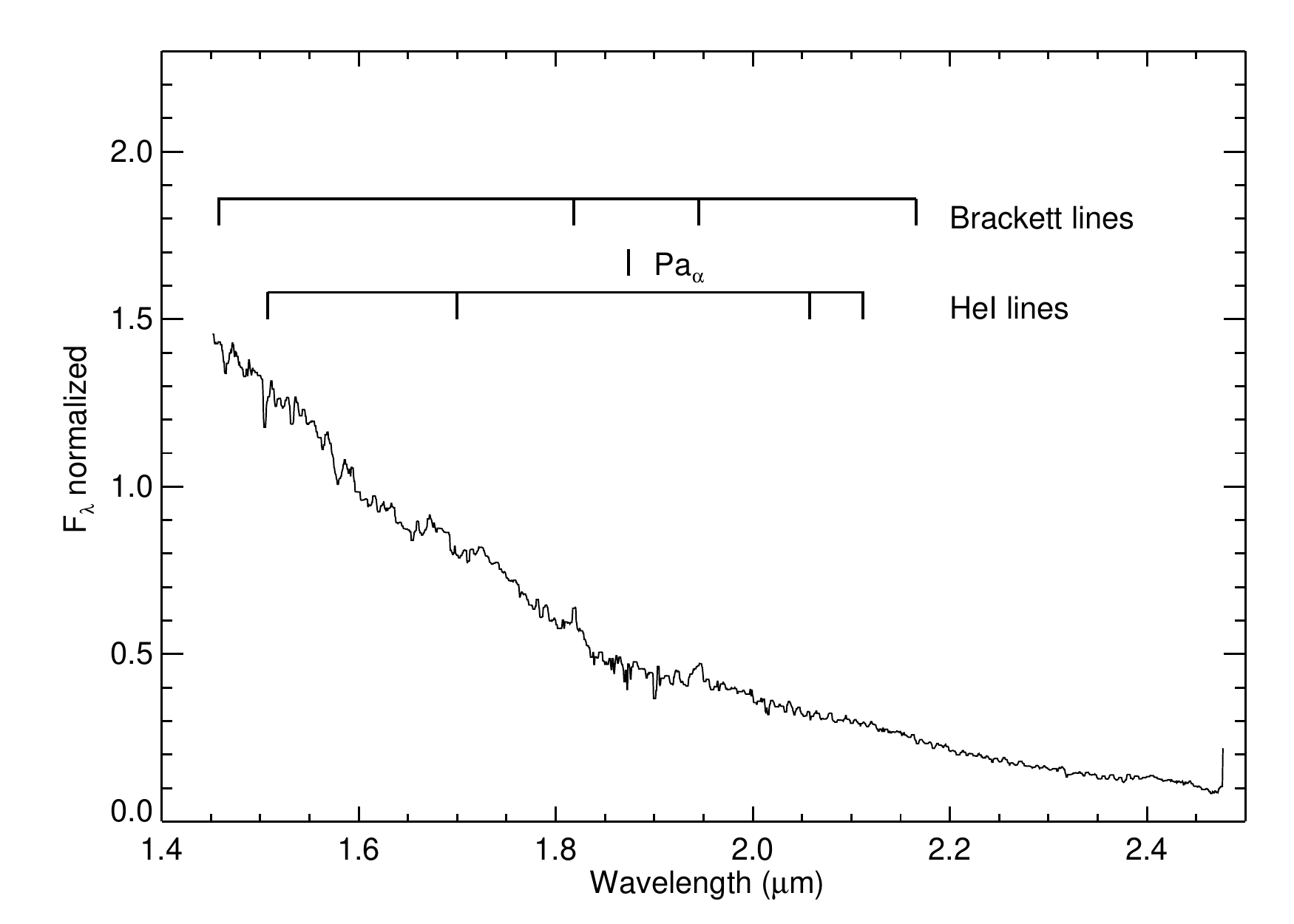}
\caption{SINFONI spectrum of HD\,8049\,B in $J$-band (left) and $HK$-band (right).}
\label{f:spectrum}
\end{center}
\end{figure*}

The companion was further characterized on August 28 UT, 2012 using the
Spectrograph for INtegral Field Observations in the Near Infrared \citep[SINFONI, 
see][]{Eisenhauer2003,Bonnet2004a,Bonnet2004b} mounted at the VLT/UT4. 
SINFONI provides AO assisted integral field spectroscopy in the 
near-infrared. It is composed by the SPectrograph for Infrared Faint Field 
Imaging (SPIFFI) fed by a modified version of the Multi-Applications Curvature 
Adaptive Optics system \citep[MACAO, see][]{Bonnet2003}. We used the gratings $J$ 
and $H+K$ to cover the 1.1--1.4 and 1.45--2.45 $\mu$m ranges at a spectral 
resolution of $\sim$2000 and $\sim$1500 respectively.  The smallest plate 
scale (12.5~mas) was used, enabling Nyquist sampling of the PSF. The primary served as a reference 
for the AO wavefront sensor and was kept outside the instrument 
field of view ($0.8~\!''\times0.8~\!''$). Each sequence was composed of 12 
acquisitions on the source with small dithering and one acquisition on the sky. 
Immediately after observing our science targets, we observed the standard star HIP\,17280 (B5V) following an AB pattern
 to correct our spectra for telluric
features. HD\,8049 and the standard were both observed under poor
atmospheric conditions with a seeing of 1.3--1.5$~\!''$, although the
AO loop remained stable, and airmass was lower than 1.1.

We homogeneously reduced our dataset with the ESO data reduction pipeline 
version 1.9.8 \citep{Abuter2006}. The pipeline successfully carried out bad 
pixels detection, flat fielding, distortion coefficient computation, and 
wavelength calibration using calibration frames acquired the day after the 
observations. Individual datacubes were reconstructed from sky-subtracted
object frames and merged into a final mosaicked datacube. Datacubes of
telluric standard stars were obtained in a similar way.
Finally, we integrated the source flux in each final datacube over
selected apertures to optimize the signal to noise ratio without
introducing differential flux losses. The spectra of the standard star
were corrected from their intrinsic features and divided by a black
body at the appropriate temperature \citep{Pickles1998}.  The SINFONI near-infrared spectrum of the object is shown in Figure \ref{f:spectrum}.

\section{Nature and properties of HD\,8049\,B}

\label{sec:companion}

\subsection{Early discovery}
\label{s:binarity}

HD\,8049 was identified as a  binary system by \citet{1929CiUO...80...59V} in 1928. 
A projected separation $\rho=3.19\arcsec$ and a position angle $\theta=153.1~ \deg$
were measured at epoch 1928.81. We corrected for the precession effect as listed 
in \citet{Heintz} and find $\theta=153.6~\deg$. The original paper did not provide 
any error bar on the measurement, just a note that there were good observing conditions. 
We assumed an error of 100 mas on $\Delta \alpha$ and $\Delta \delta$. The magnitude 
resulting from the visual observations was 13.5~mag (rounded to 0.5~mag) and a bluish 
color was reported; the assumed error {is} $0.5$~mag. This is the only available 
measurement of the pair listed in the Washington Double Star (WDS) Catalog \citep{wds}
before our observations. This position is possibly compatible with the one measured in our EFOSC $U$-band {images} and with {the} NACO results, once the time lapse is considered. 
Table \ref{t:astrometry} summarizes the available astrometry, including the two epochs 
from NACO.  

From NACO observations spanning over 1~yr, we derive a relative motion of 2.7~mas/yr in
$\alpha$ and 30.4~mas/yr in $\delta$.  Doing the same for 1928 vs 2010
observations we get a motion of 0.4~mas/yr in $\alpha$ and 26~mas/yr
in $\delta$.  These determinations are obviously much smaller than the
proper motion of the star ($\mu_{\alpha} = 65.99\pm1.18$~mas/yr and
$\mu_{\delta} = -240.99\pm0.98$~mas/yr), supporting the physical
association of the faint companion. The motions
with respect to the central star derived from 1928 to 2010 and from 2010
to 2011 are consistent within their errors, making viable the
hypothesis that the companion detected in 1928 in the optical range is the
same {one} we found with NACO, and that we are indeed tracking the orbital motion of
HD\,8049\,B around HD\,8049\,A. The characterization of the orbital
properties of the binary is discussed in Section \ref{sub:orbit}.

\begin{table}[h]
\caption{Astrometry of candidate(s) around HD\,8049}
\label{t:astrometry}
\begin{center}       
\begin{tabular}{l l l l} 
\hline\hline
Epoch             &  $\Delta \alpha$ &  $\Delta \delta$  &  Reference \\
                  &   (mas)            &   (mas)             &  \\
\hline 
1928.81             &  $1417\pm100$     &   $-2858\pm100$     &  \citet{1929CiUO...80...59V}\\
2010.58      &  $1378\pm7$   &   $-744\pm7$ & NACO \\
2011.58      &  $1382\pm10$   &   $-705\pm10$    & NACO \\
2011.97  &   $1373\pm24$  &   $-734\pm24$     & EFOSC \\
\hline 
\end{tabular}
\end{center}
\end{table}

\subsection{HD\,8049\,B, a new nearby WD}
\label{sec:wd} 

Considering the combined optical and NIR photometry of HD\,8049\,B ($U$, $V$
and $H$ bands) from EFOSC, the WDS Catalog and NACO observations, the
hypothesis of a sub-stellar companion is unambiguously excluded. The
combined visible and NIR photometry indicate HD\,8049\,B
is {most likely} a WD. The SINFONI spectrum indeed confirms this assumption as the 
observed spectral energy distribution shows a hot and dense atmosphere that 
can be associated {with} that of a WD. The NIR spectrum is rather 
featureless as we do not detect any hydrogen or helium lines over the whole
1.0--2.5$~\mu$m spectral range. Despite the low-SNR, it could indicate
a featureless Helium-rich nature (DB WD) or a Hydrogen-rich white 
dwarf (DA) with a relatively hot atmosphere and therefore less contrasted 
Hydrogen lines \citep[see][]{Dobbie2005}. Visible band spectroscopy should enable 
a better characterization of the WD nature of HD\,8049\,B, but 
unfortunately the low quality EFOSC spectrum does not provide such additional 
information due to the blend with the much brighter primary spectrum.

To further constrain the physical properties of HD\,8049\,B, we used the
emitted fluxes in the $FUV$ and $NUV$, listed in the GALEX Catalog 
\citep{2005ApJ...619L...1M}. In the ultraviolet, we applied the correction 
for the chromospheric emission of the primary as described in Sect.~\ref{sub:photostar}. 
We refer to that section for an exhaustive summary of the photometry for the two objects.

As a preliminary analysis of the object properties, we created empirical sequences using the 
catalog of nearby WD{s} by \citet{Giammichele2012}. {When required photometry was unavailable there,}  supplemented it with available 
GALEX $FUV$ and $NUV$ magnitudes, and 2MASS $J$, $H$ and $K_S$ magnitudes. The final sample consisted of 107 nearby ($\leq 51$~pc) 
WDs: 22 with $FUV$ magnitude{s}, 18 with $NUV$ magnitude{s} and 84 with 
$J$, $H$, $K_S$ magnitudes calibrated by \citet{Giammichele2012}. 

\begin{figure*}
\centering
\includegraphics[width=0.4\textwidth]{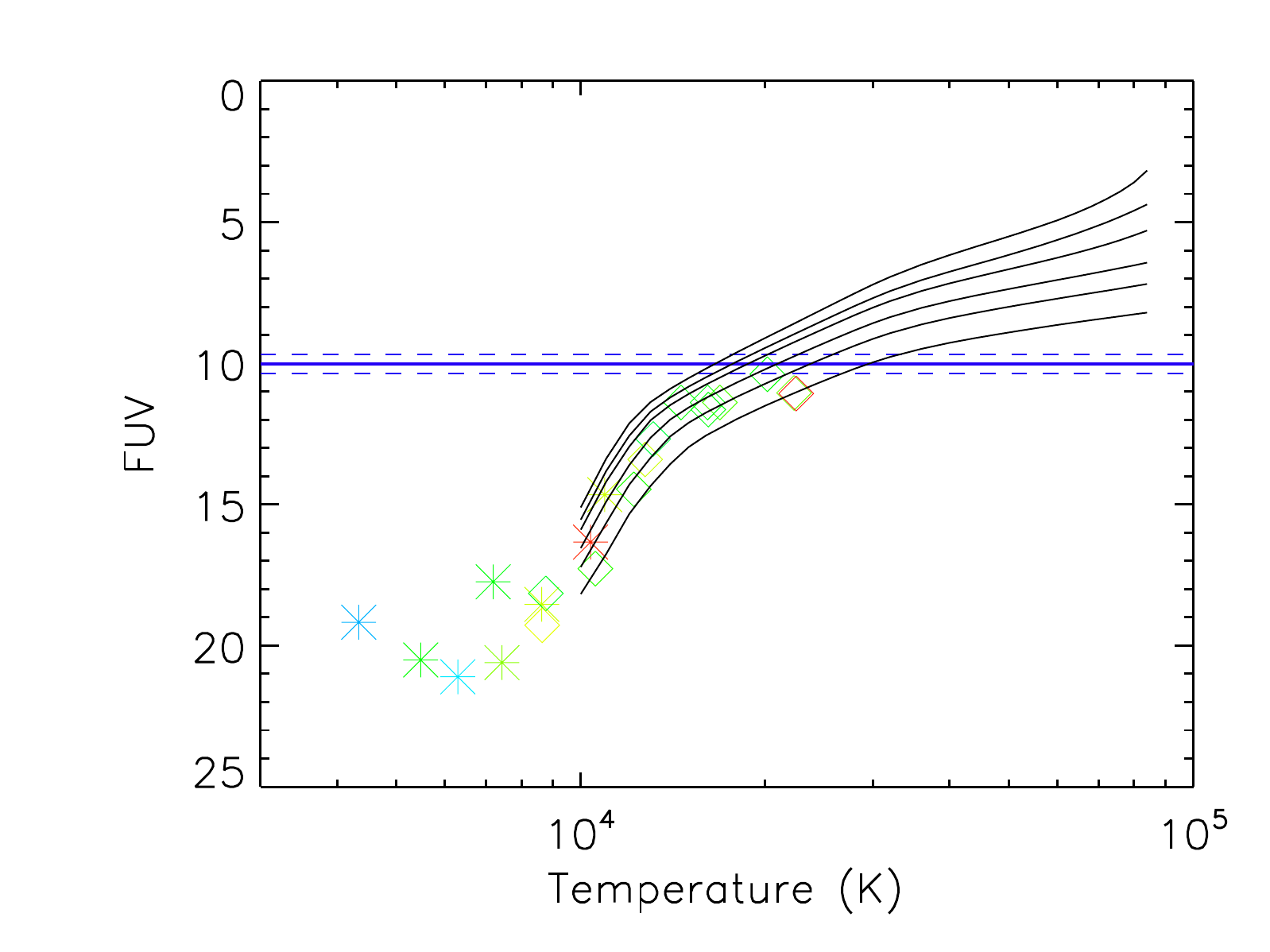}
\includegraphics[width=0.4\textwidth]{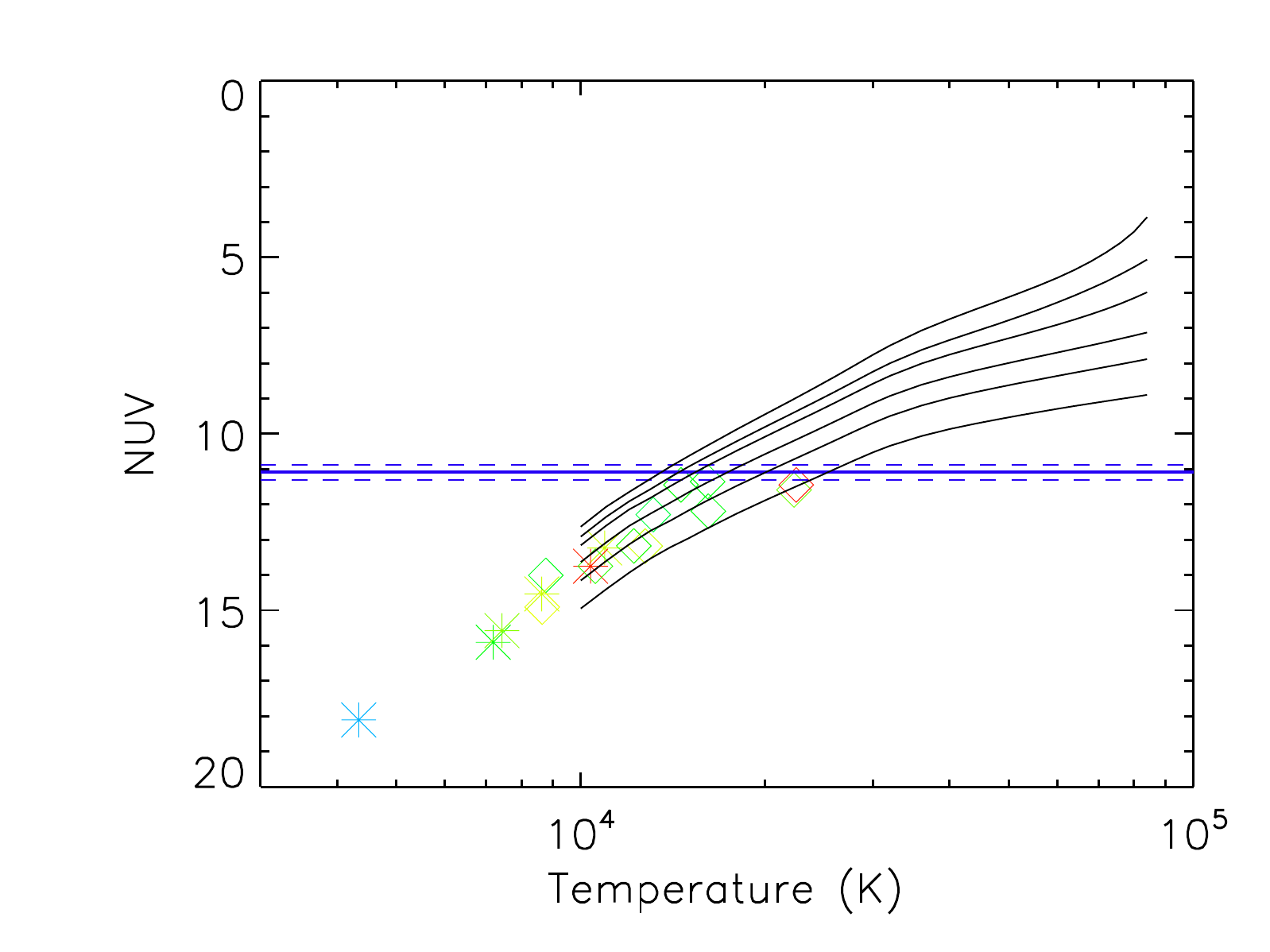}

\includegraphics[width=0.4\textwidth]{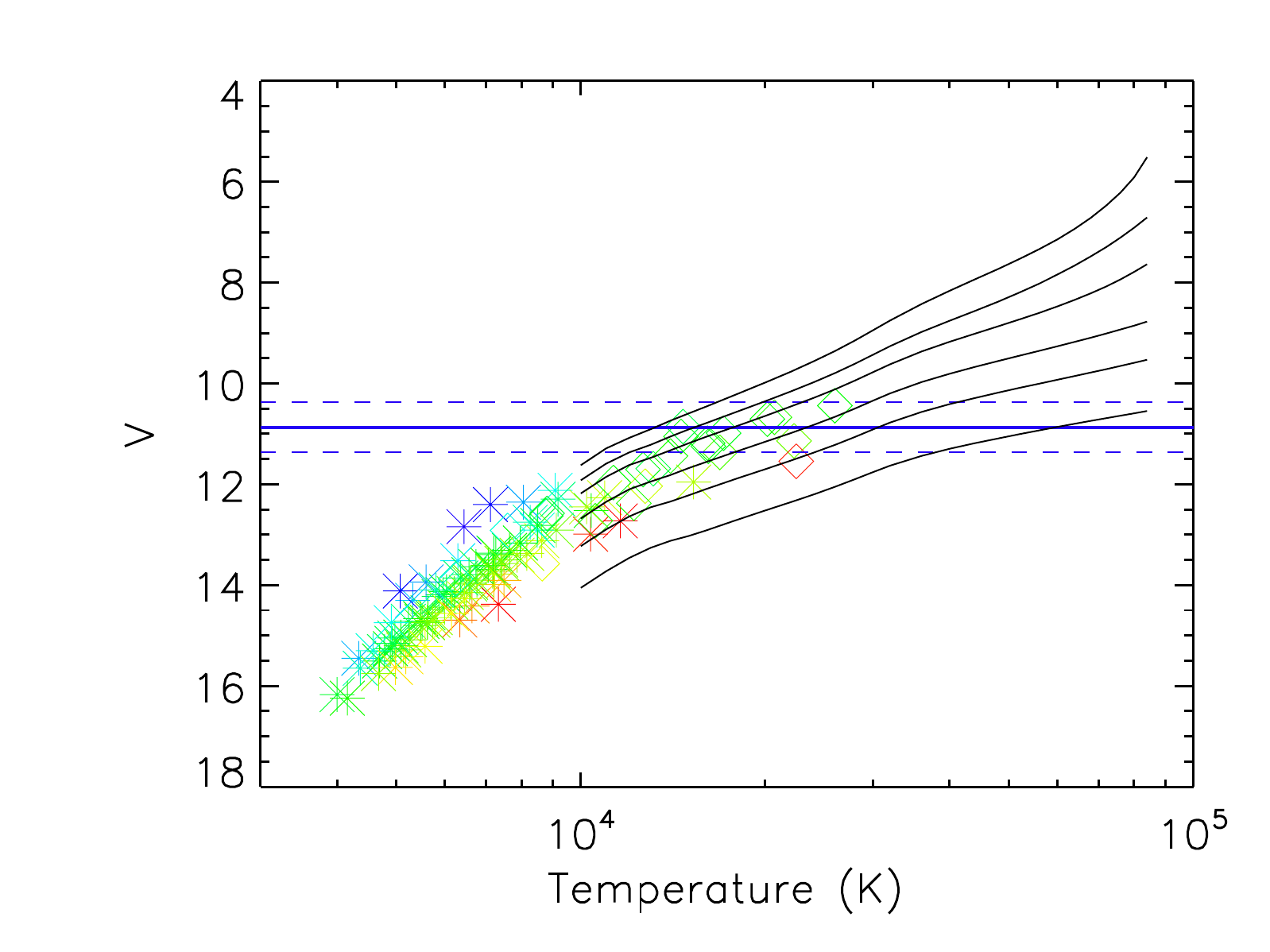}
\includegraphics[width=0.4\textwidth]{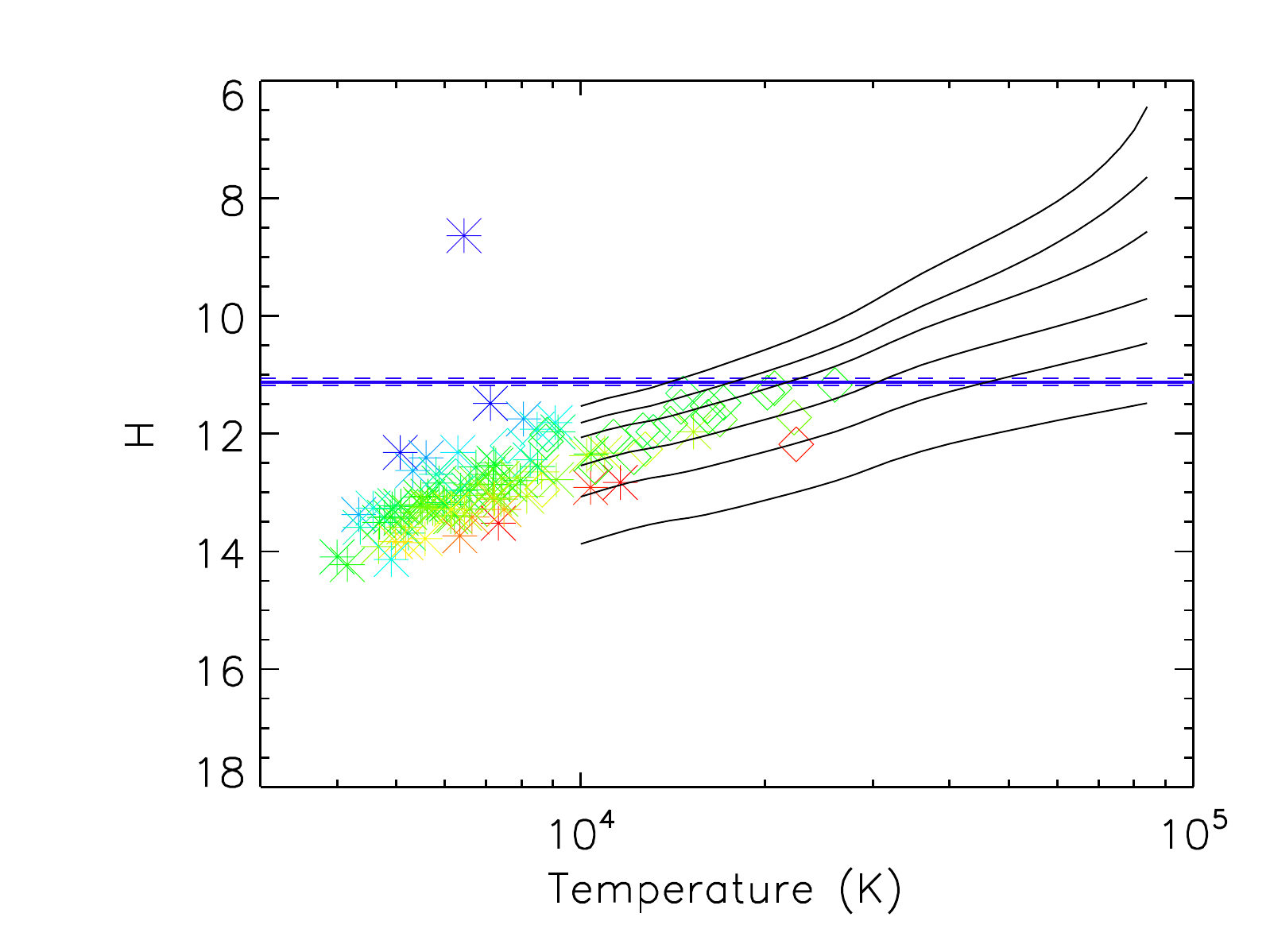}
\caption{{Absolute magnitude in different bands (from top-left $FUV$, $NUV$, $V$ and $H$) versus effective temperature for white dwarf models of \citet{Ven11} (black lines) and sample a of nearby dwarfs collected by \citet{Giammichele2012}.  Stars represent objects with all magnitudes available and diamonds represent the others. Colors indicate the mass from the lowest (blue) to the highest (red ones). The continuous blue horizontal line represents the magnitude of HD\,8049\,B. The corresponding error bars are plotted as dashed blue lines. The plots show that the objects of the \citet{Giammichele2012} catalog are not peculiar and are well described by the theoretical models of \citet{Ven11}.} }
\label{f:wdtemp}
\end{figure*}

The catalog of 
observed WDs listed mass, $T_{\mathrm{eff}}$, $\log g$ (logarithm of the surface gravity) and $t_{\mathrm{cool}}$, all measured 
from spectroscopy.
As a check, we overplotted these object parameters with the expected values from 
theoretical sequences of \citet{Ven11}, and confirmed that they perfectly fit, as shown in Figure \ref{f:wdtemp}.

We noticed that the $FUV - NUV$ color of HD\,8049\,B did not fit with theoretical nor empirical models.  The $NUV$ magnitude is more affected by the chromosphere of the K star, but if we considered the corrected value, the discrepancies were even greater.  
 
In order to fit our photometry measurements with the SED (Spectral Energy Distribution) of a white dwarf object, we used the MCMC (Markov Chain Monte Carlo) based code PASTIS (D\'iaz et al., in preparation). 
To extract characteristics of the WD atmosphere, we used models by \citet{Koester2010}; these models predict
the mean intensity of the stellar disk and depend on $\log g$, $T_{\mathrm{eff}}$ and the 
atmosphere composition; the latter is assumed to be of pure H in this case. To calculate 
the observed flux measured from our viewpoint on Earth, we multiplied the intensity by the solid angle of the star, calculated using the distance and the radius of the WD. The latter was obtained from the $T_{\mathrm{eff}}$--$\log g$--$t_{\mathrm{cool}}$--mass relation from \citet{2006AJ....132.1221H,2006ApJ...651L.137K,2011ApJ...730..128T,2011ApJ...737...28B}\footnote{http://www.astro.umontreal.ca/$\sim$bergeron/CoolingModels} (used later on to also compute the derived parameters).

The synthetic spectrum was then corrected for interstellar extinction using the relations of 
\citet{Fiz1999} with $R_{V}$ = 3.1, and an assumed reddening value of $E_{\mathrm{(B-V)}} = 
0.0106$, as listed in the GALEX catalog.
Finally, the spectrum was integrated in the bandpass listed in Table~\ref{t:photom} as described in 
\citet{Bayo2008}. We did not consider the $NUV$ point for the reason listed {previously.  We also omitted} the $U$ point because {it} can be affected by the cut-off of the atmosphere; the $V$ point was rejected because the observation was made by eye and there is not an associated error. 

\begin{figure}[!ht]
\centering
\includegraphics[height=5.5cm]{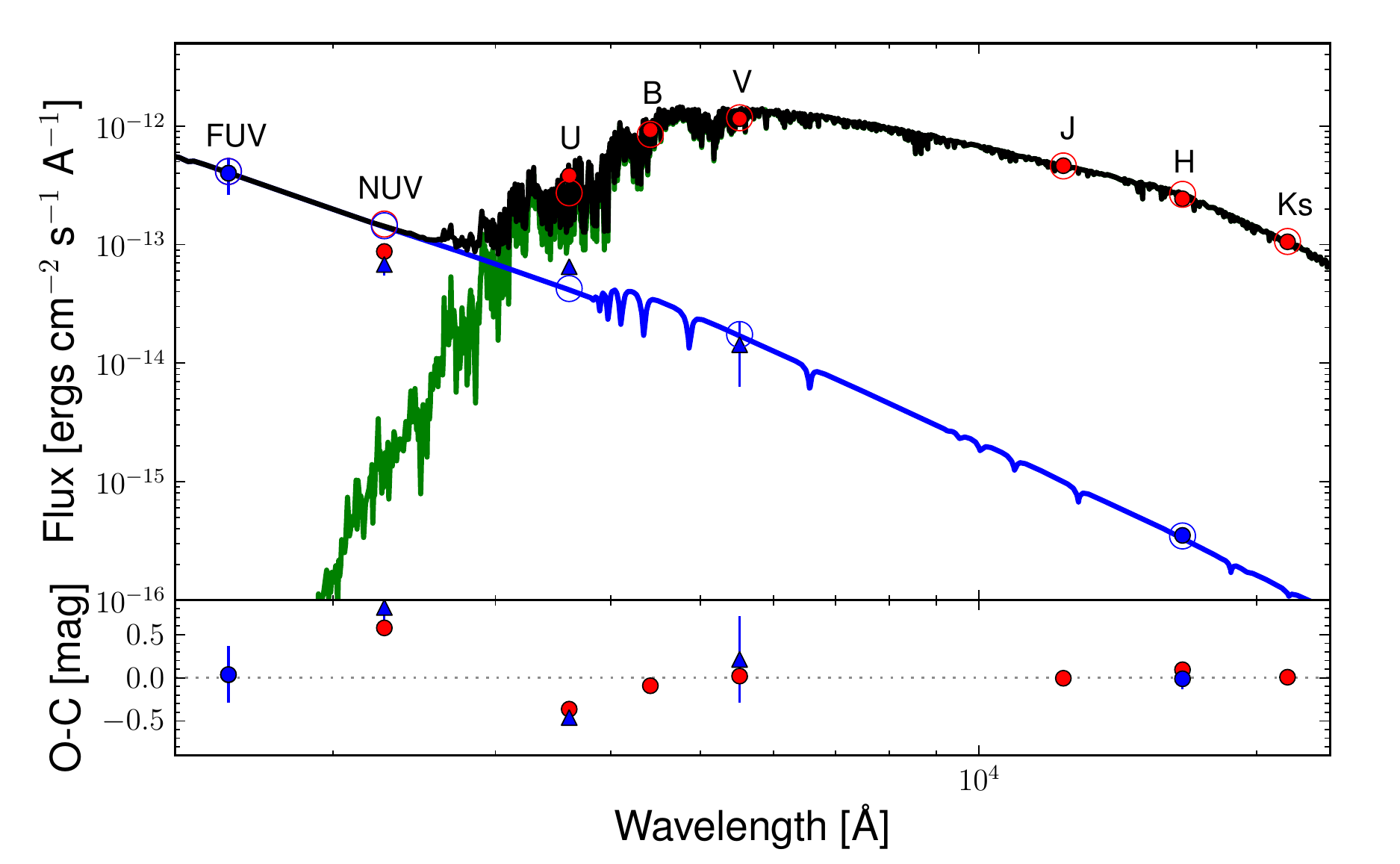}\vspace{1cm}
\includegraphics[height=8.2cm]{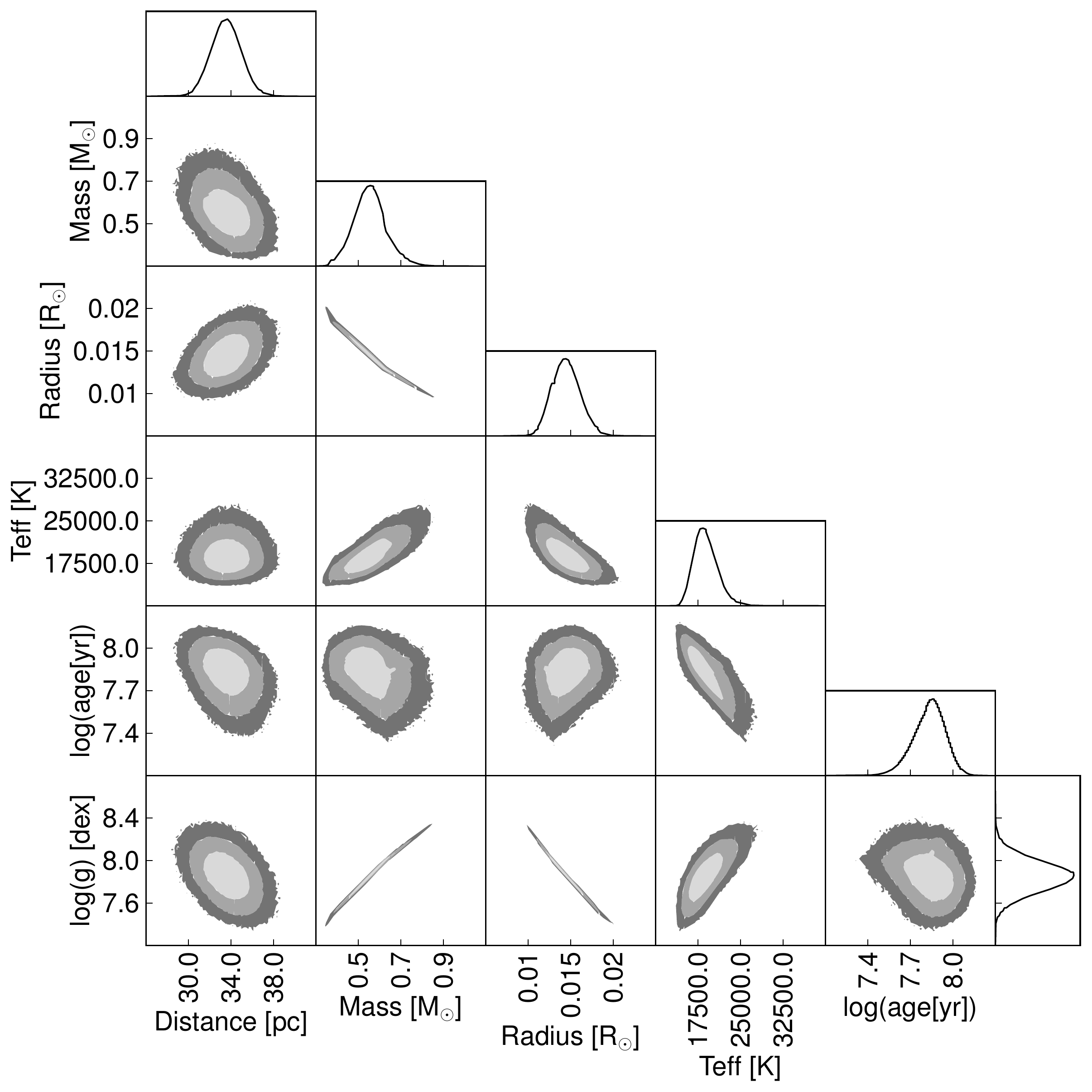}
\caption{ Top: Observed magnitudes of the system plotted over the synthetic spectra. Blue points are those taken in consideration during the SED fitting of the WD. Triangle shaped ones are not considered during the PASTIS integration. Values of the magnitudes of the integrated system are represented in red and the SED in black. The primary star SED is represented in green.  Bottom: Two parameters joint posterior distributions for HD\,8049\,B. The 68.3\%, 95.5\% 
and 99.7\% confidence regions are denoted by light grey, dark gray and black areas 
respectively. Single PDF (Posterior Distribution Function) of each combination is shown along the sides.}
\label{f:sed}
\end{figure}

The SEDs of the system and its components are shown in Figure~\ref{f:sed}.  The fitted parameters are the distance, $\log g$ and $T_{\mathrm{eff}}$. As priors, we assumed a normal distribution 
with a mean distance of $33.6$~pc and a standard deviation of $1.4$~pc (based on the Hipparcos parallax of $29.79\pm 
1.21$~mas, \cite{2007A&A...474..653V}; for $\log g$ we used a uniform distribution in the range 7.0--9.0; for $T_{\mathrm{eff}}$ we used a uniform distribution in the range 10000--40000~K.
 
We ran 50 chains of $5\times 10^5$ steps starting from random values of the priors. After removal of the {\it burning phase} (steps before convergence), we merged all the chains, taking into account the correlation lengths. The results of the fitted parameters are $\log g = 7.86 \pm 0.16$ and $T_{\mathrm{eff}} = 18800 \pm 2100$~K. The distance remained equal to the prior. The derived parameters were $M_{\mathrm{WD}} = 0.56 \pm 0.08~M_{\odot}$, 
$R_{\mathrm{WD}} =  0.0145 \pm 0.0016~R_{\odot}$, and cooling time $t_{\mathrm{cool}} = 63^{+23}_{-17}$ Myr (see Figure~\ref{f:sed}). Note that the errors of the values are the 68.3\% confidence interval of the fit, and they do not take into account possible model errors.

\begin{figure*}
\centering
\includegraphics[width=0.4\textwidth]{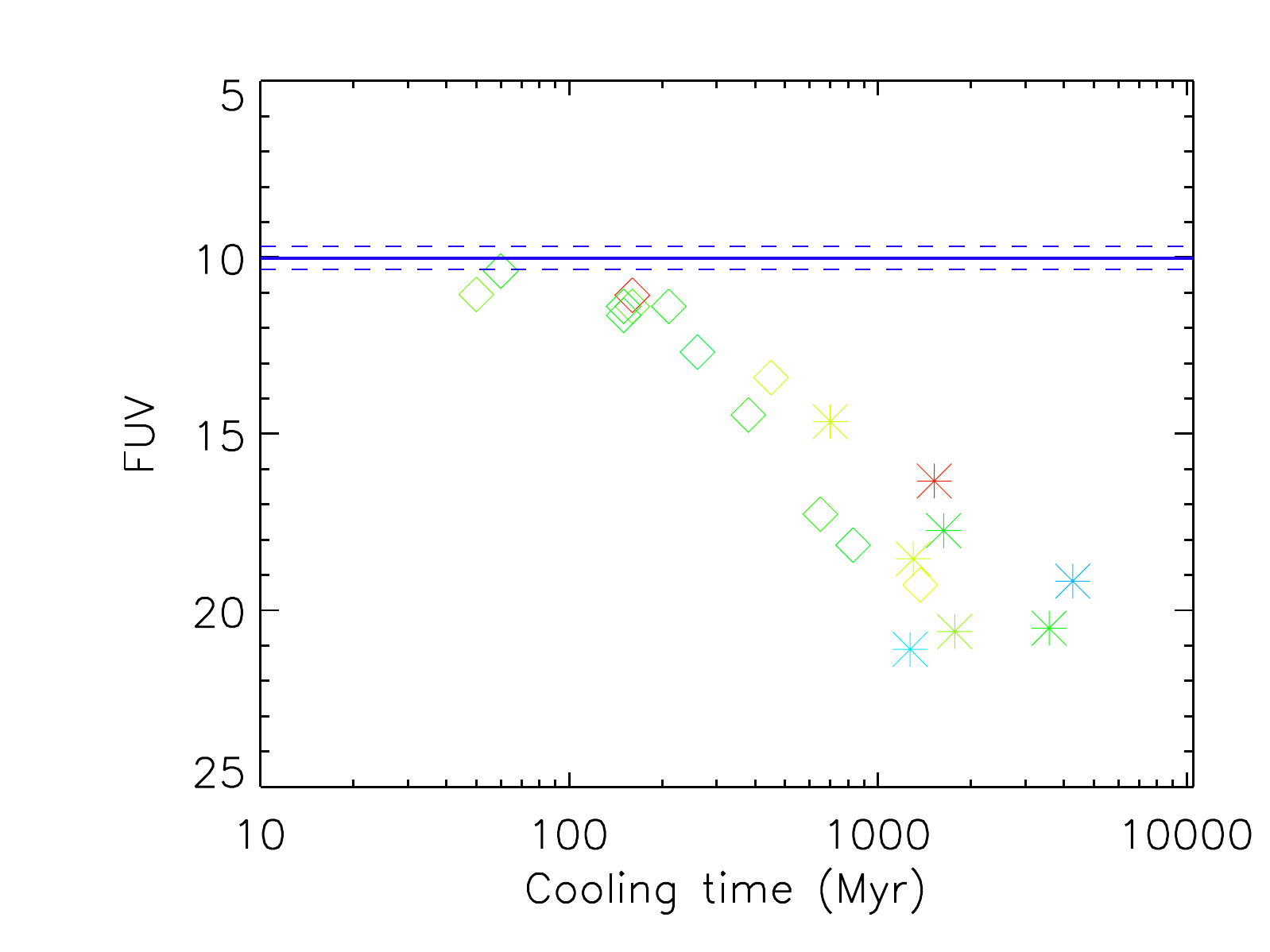}
\includegraphics[width=0.4\textwidth]{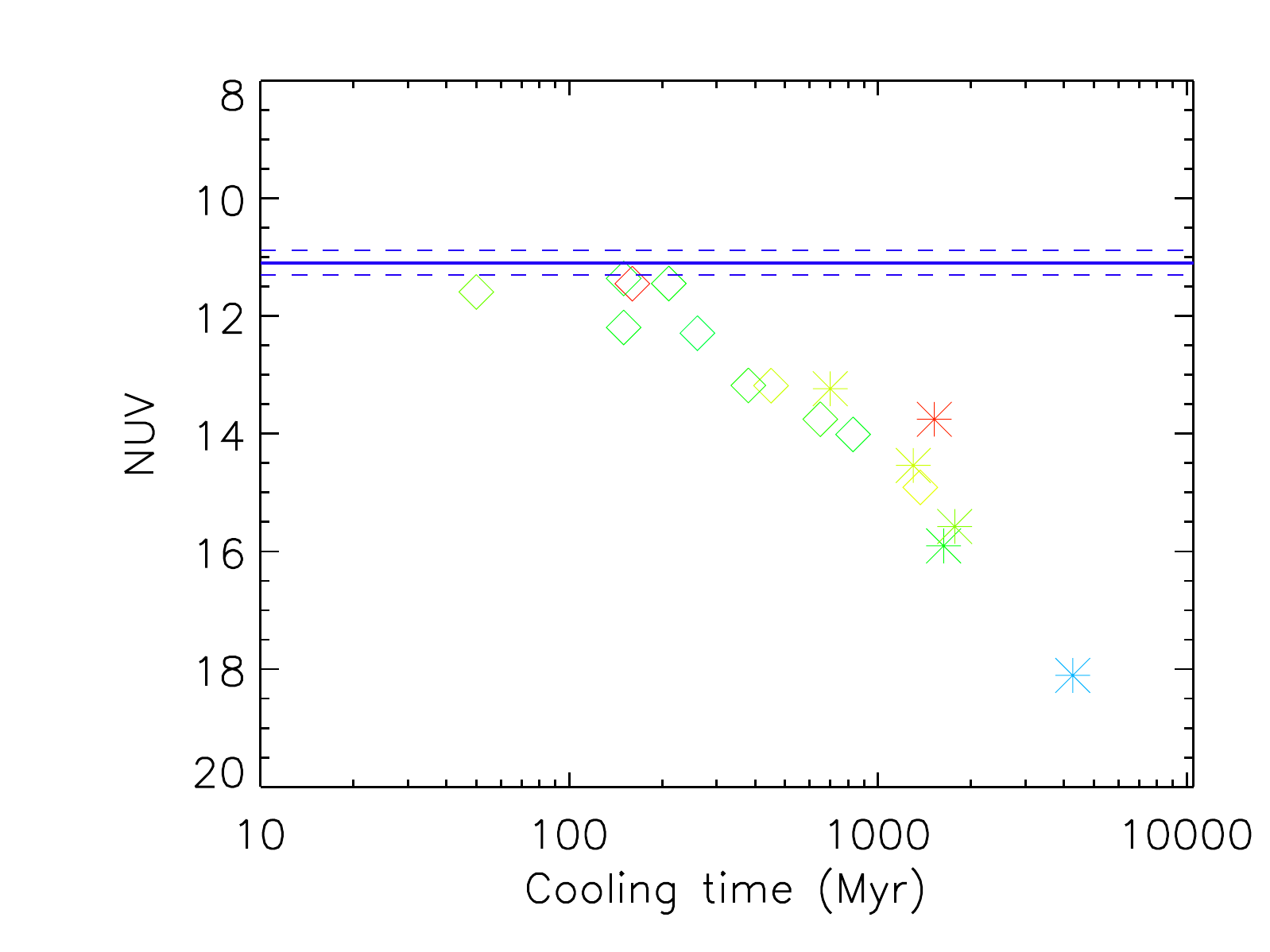}

\includegraphics[width=0.4\textwidth]{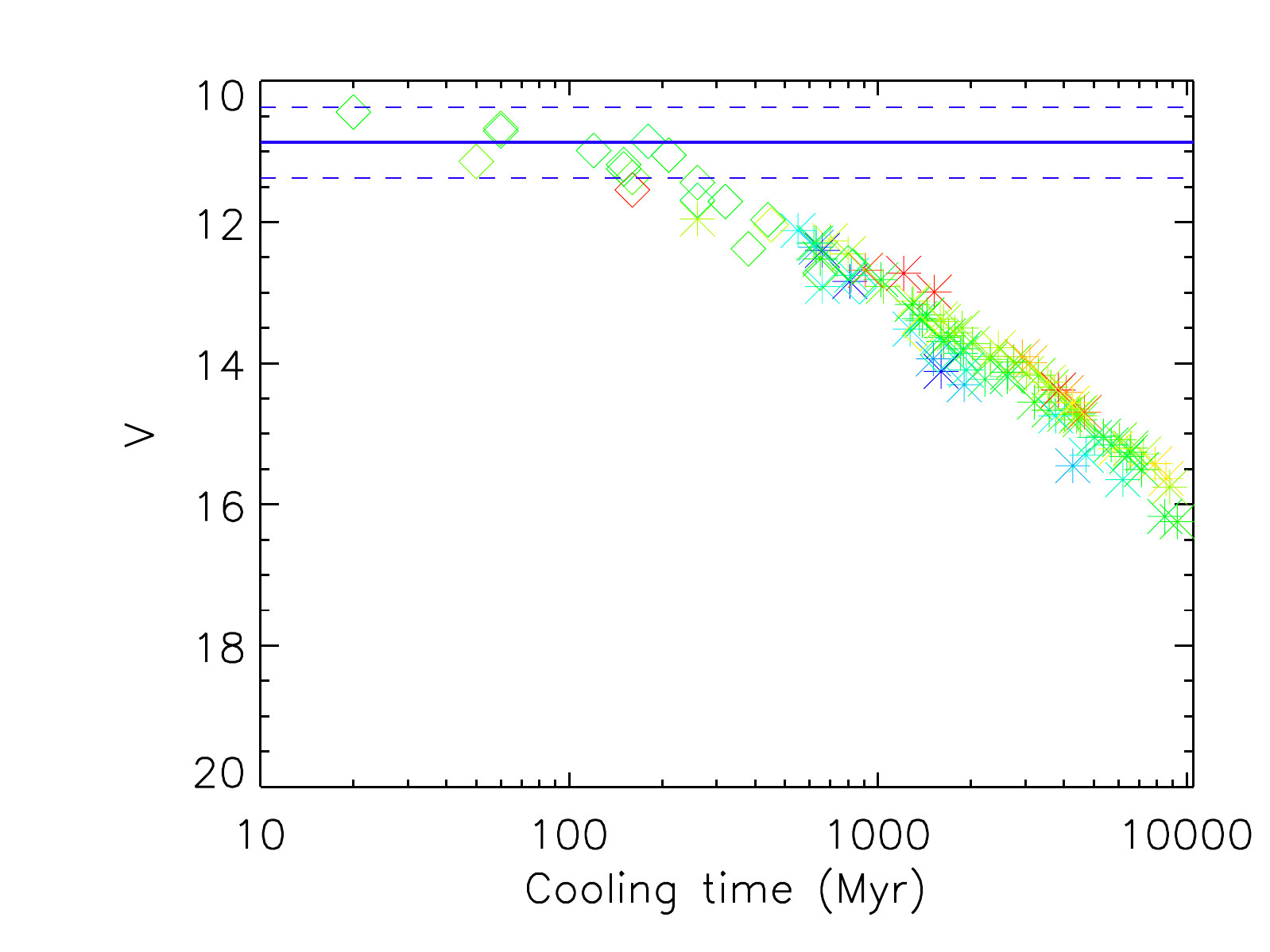}
\includegraphics[width=0.4\textwidth]{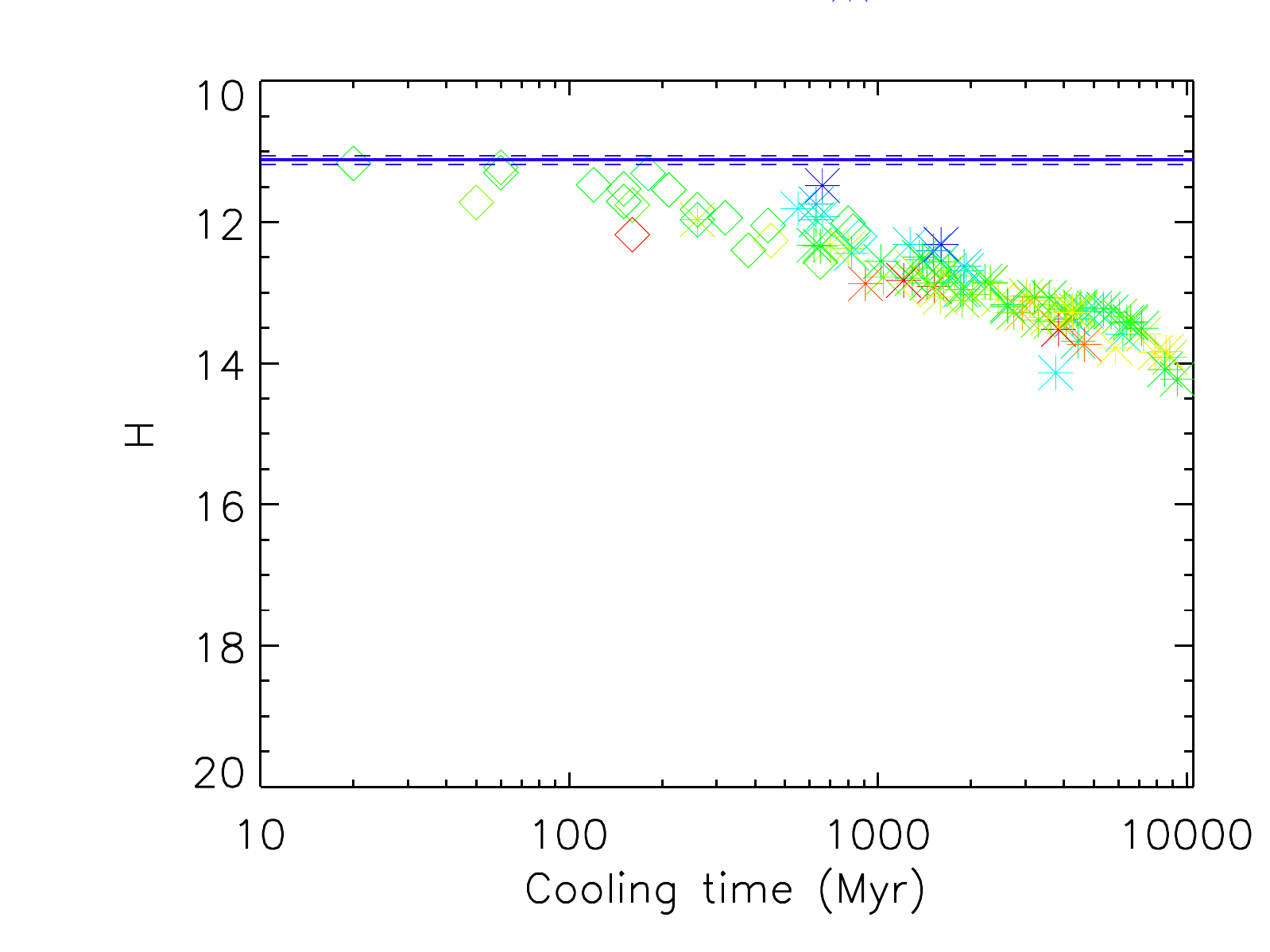}
\caption{{Absolute magnitude in different bands (from top-left $FUV$, $NUV$, $V$ and $H$) versus cooling age for white dwarf models of \citet{Ven11} (black lines) and sample a of nearby dwarfs collected by \citet{Giammichele2012}.  Stars represent objects with all magnitudes available and diamonds represent the others. Colors indicate the mass from the lowest (blue) to the highest (red ones). The continuous blue horizontal line represents the magnitude of HD\,8049\,B. The corresponding error bars are plotted as dashed blue lines. The plots show that the objects of the \citet{Giammichele2012} catalog are not peculiar and are well described by the theoretical models of \citet{Ven11}.}}
\label{f:wdage}
\end{figure*}

The cooling time calculated using the empirical sequences is $98_{-40}^{+28}$~Myr 
(see Fig.~\ref{f:wdage}), in agreement with the SED fit.  {To estimate the age of the system we must add the main sequence lifetime 
to the cooling age of the WD. This is discussed in Sec.~\ref{sub:history}.}

\section{Nature and properties of HD\,8049\,A}
\label{sec:star}
We exploited the spectroscopic data described in Sec. \ref{sub:spectra} and 
additional data from public archives with the aim of improving 
the age determination and other stellar parameters of the 
primary. 

\subsection{Rotation period}

HD\,8049 was observed in the All Sky Automated Survey
\citep[ASAS\footnote{http://www.astrouw.edu.pl/asas/} Variable star
catalog;][]{Pojmaski2002}. Archive observations span from November 20 UT, 2000
to October 28 UT, 2009. After removing outliers and averaging observations 
collected within short time intervals (less than 2~hr), we were left with 583 
data points for the rotation period search.  The average photometric accuracy 
was 0.02~mag. The brightest $V$ magnitude was $V = 8.68 \pm 0.02$~mag. To
minimize the effects of active region growth and decay on the periodograms, 
we sectioned the data time series into 14 light curves. The light curve 
amplitude was variable, with values up to $\Delta V = 0.04$ mag. The 
Lomb-Scargle analysis \citep{Scargle1982} was carried out on the complete 
time series data as well as on each of the 14 light curves. We found a 
rotation period of $P  = 8.3\pm0.1$ days in the complete time series data, 
and the same period in 4 out of 14 time intervals (with FAP, False Alarm 
Probability, $\le$ 1\%). {The light curve amplitude was larger than the photometric 
accuracy $\sigma$ in 13 out of 14 time segments.}

\begin{figure}
\centering
\includegraphics[height=9.5cm]{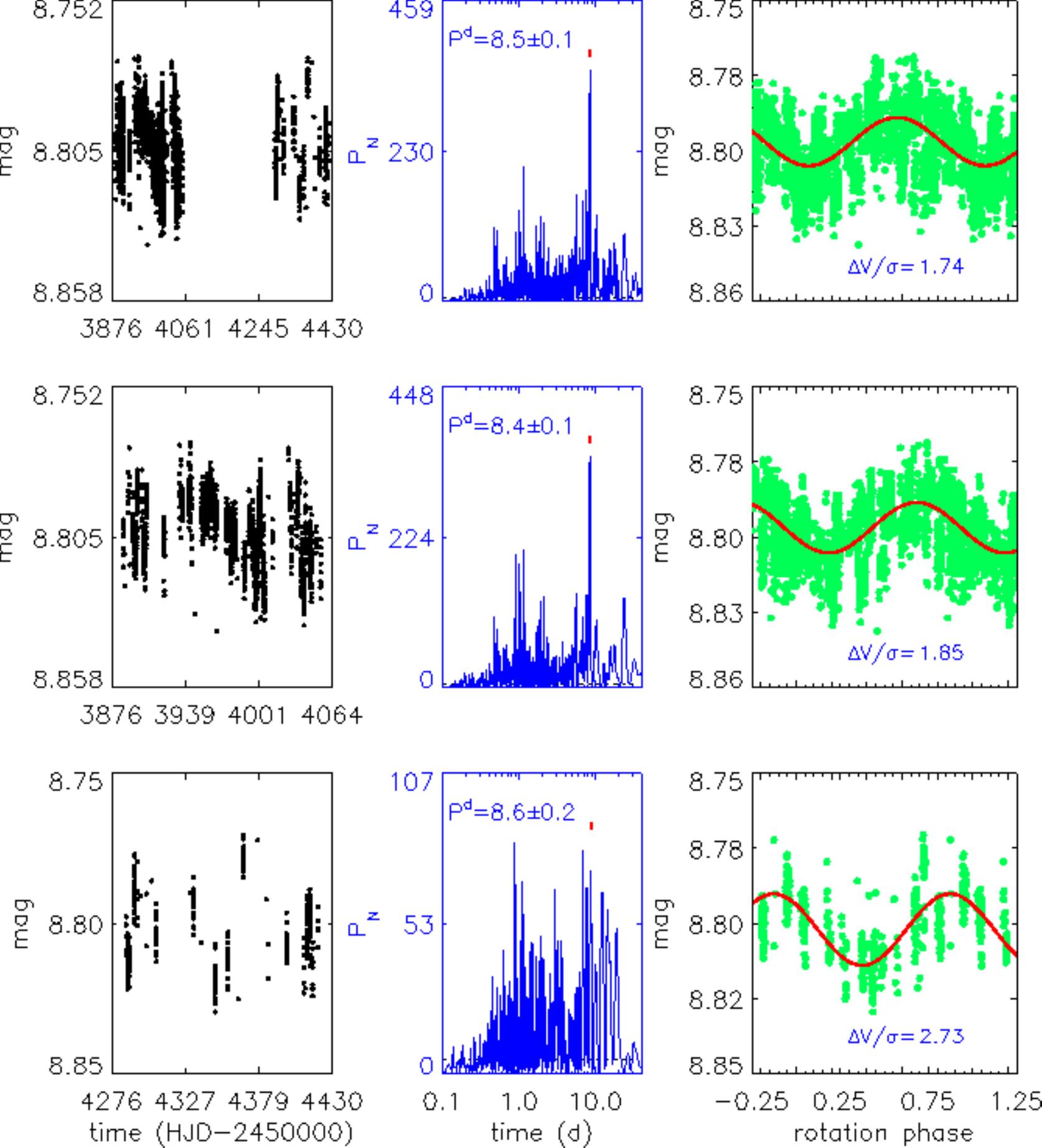}
\caption{Photometry time series of HD\,8049 from SuperWASP: left columns display time 
segments of magnitudes versus Heliocentric Julian Days {(HJD)} (the first panel only shows 
the complete series). Middle columns display the Lomb-Scargle periodograms with 
indication of the rotation period. Right columns display the light curves phased 
with the rotation period and the first HJD as initial epoch. Solid {curves} represent 
the sinusoidal fit with the rotation period. The labels report the ratio of average 
residual from the fit and the rotational modulation amplitude.}
\label{f:prot}
\end{figure}

HD\,8049 was also observed by SuperWASP \citep{Pollacco2006, Butters2010}. 
The public archive permitted us to retrieve data for the 2006 and 2007 seasons\footnote{http://www.wasp.le.ac.uk/public/}. We carried out an analysis 
similar to that described for the ASAS data. After outlier removal, averaging,
and data series sectioning, we found a rotation period $P = 8.5\pm 0.1$~days for 
the complete series as well as for both 2006 and 2007 seasons, with a confidence 
level larger than 99\%. {This is consistent} with the period determination from 
ASAS data. The maximum light curve amplitude was 0.02~mag, whereas 
photometric accuracy was 0.01~mag. In Fig.~\ref{f:prot} we show a summary of the 
period search on Super WASP data. We refer the reader to \citet{Messina2010} for 
a more detailed description of the photometric data analysis methods. According to 
the criteria adopted in \citet{Mes11} for assigning a confidence level to the rotation 
period, the period of HD\,8049 can be considered {\it confirmed}. 
The $P$ = 8.3 days rotation period is compatible with the projected rotational 
velocity \citep[$3.5\pm1.2$~km/s, see][]{Tor06} for stellar radii 
R $\ge$ 0.6~R$_{\odot}$.

\subsection{Age indicators}
\label{sub:age}

Table \ref{t:age} summarizes the measurements and corresponding ages as
derived from {the} different indicators.  From our analysis of the HARPS
spectrum, we confirm that no lithium is detectable, as found by
\cite{Tor06}.  Two sets of $\log R_{\mathrm{HK}}$ measurements
are available: those from our CORALIE spectra and that by
\citet{Gray2006}. {Within the uncertainties they give} the same value ($\log
R_{\mathrm{HK}}=-4.25$). {The logarithm of the ratio of X--ray and bolometric luminosity} $\log L_{\mathrm{X}}/L_{\mathrm{bol}}$ was derived
from ROSAT All-Sky Bright Source Catalog \citep{1999A&A...349..389V}
assuming the source \object{1RXS J011915.1-433803} is the X--ray
counterpart of HD\,8049\,A.  The X--ray emission should be dominated by the
active K dwarf primary since HD\,8049\,B is cooler than WDs that show
significant X--ray emission \citep[$T_{\mathrm{eff}}>25000$~K,
][]{1999ApJ...525..995V}. {The $U$, $V$ and $W$ heliocentric space velocities (given in a right handed coordinate 
system with U towards the galactic centre) are taken from 
\citet{Holmberg} adopting the trigonometric parallax of $ 29.79 \pm 1.21$~mas \citep{2007A&A...474..653V}}. The
uncertainty in the center of mass velocity due to orbital motion spans 
from a minimum value of 200~m/s to a maximum of 3~km/s, as determined from the simulations 
described in Sec.~\ref{sub:orbit}. 

The age estimates provided from these various indicators are quite discordant.
Chromospheric HK emission, coronal emission, and rotation period
indicate an age between 100--400~Myr. On the other hand, the lack of
lithium indicates an age older than about 500~Myr. {Additionally} the kinematic
parameters are quite far from the typical locus of young stars
{\citep{Montes2001}}. {The galactic orbit as given in
\citet{Holmberg} has $e = 0.19$ and
$z_{max} = 0.33$~kpc. These properties are more typical of stars with age of about 3 Gyr,
as estimated from the galactic orbits of stars with small age uncertainty
in the \citet{Holmberg} catalog.}  These discrepancies strongly suggest that HD\,8049\,A is not a bona-fide
young star but is rather a peculiar object. 

We identify the source of these anomalies {to be} the WD 
companion detected through direct imaging. The accretion of some amounts of 
mass and angular momentum by the central star at the time of mass loss from 
the WD progenitor might have spun-up the star, while increasing 
the level of magnetic activity. The presence of the WD {cannot} alter 
significantly the measurement of the rotation period and  X--ray flux. 
{O}nly minor effects on {the} $\log R_{\mathrm{HK}}$ measurement {are} expected 
from the contamination of the spectrum. 

We conclude that the young rotation-activity age is not spurious.
The central star should have been rejuvenated by the accretion of some
amount of mass and angular momentum at the time of
mass loss from the WD progenitor. This scenario {is}
investigated {more thoroughly} in Sec.~\ref{sub:history}.

\begin{center}   
\begin{table}
\caption[]{Age indicators of HD\,8049\,A. The 2nd column lists the adopted value for
the various parameters, and the 3rd column lists the corresponding reference; the 4th column contains
the resulting age, and the last column has the adopted calibration for age determination.} 
\label{t:age}
\begin{tabular}{lllll}
\hline
\hline
Indicator                  &  Measure    & Ref &   Age (Myr)      &  Ref. \\
\hline
Li EW   (mA)               &  0       & 1,2 & $>$500     &  1 \\
$\log R_{\mathrm{HK}}$              &  $-4.25\pm0.05$      & 1,3 &     90     &  4 \\
$\log L_{\mathrm{X}}/L_{\mathrm{bol}}$       &    {-4.24}        & 1   &    182     &  4 \\
$ P_{\mathrm{rot}}$ (d)             & $8.3\pm0.1$ & 1   &    360     &  4 \\
$ P_{\mathrm{rot}}$ (d)             &             &     & $380\pm30$ &  5 \\
$U$, $V$, $W$ (km/s)   & 18,$-47$,$-28$  &  6   & old (few Gyr) & 1 \\

\hline

\end{tabular}
\tablebib{1: this paper;
 2: \citet{Tor06};
 3: \citet{Gray2006};
 4: \citet{Mam08};
 5: \citet{Delorme2011}
 6: {\cite{Holmberg}}}
\end{table}
\end{center}

\subsection{Mass determination}
\label{sub:mass}
Using our spectroscopic temperature 
and metallicity results, we employ the PARAM interface\footnote{\tt http://stev.oapd.inaf.it/param} 
\citep{2006A&A...458..609D} to conclude a mass of $0.804 \pm 0.023~M_{\odot}$. The mass of the 
convective envelope of the star, following \citet{2001ApJ...555..801M}, is $0.086~M_{\odot}$.

\subsection{Abundance analysis}
\label{sub:abu}

{It is well known that a considerable amount of mass can be transferred to a 
companion in close binary systems. This is the explanation for blue 
stragglers in clusters \citep{McCrea1964}, Ba-stars and extrinsic 
C-stars \citep{McClure1980, McClure1984, McClure1990}. Different signatures can be observed, depending on when 
and how much mass was transferred from the donor (in our case the white dwarf). The surface chemical 
composition of the current primary can therefore be used to better constrain the 
history of this system.}

In order to better characterize the primary, and to explore possible
signatures of accretion of material processed by the WD progenitor, we
derived the stellar spectroscopic parameters and measured the abundances
of several elements. To this aim, we used the HARPS high resolution
spectrum of HD\,8049\,A (Sec.~\ref{sub:spectra}) and adopted the
prescriptions of \citet{Biazzo2012}. For a detailed description 
of the method, we refer to that paper.

The derivation of the iron abundance of our target was performed differentially
with respect to the Sun. To this purpose, we analyzed a Ganymede spectrum 
acquired with HARPS at high $S/N$, obtaining $\log n$(FeI)$_\odot$ = $\log 
n$(FeII)$_\odot$ = $7.53\pm0.05$. In the end, we found the following stellar 
parameters and iron abundance for HD\,8049\,A: effective temperature 
$T_{\rm eff} = 5050\pm50$~K, surface gravity $\log g = 4.50\pm0.10$~dex,
microturbulence velocity $\xi$ =$ 1.42\pm0.05$~km/s, [FeI/H] = $-0.10\pm0.02$ dex, 
and [FeII/H] = $-0.10\pm0.05$ dex. 

Looking for potential evidence of pollution by the ejecta of the precursor of 
the WD, we derived abundances of the s-process elements\footnote{Elements heavier than iron are mostly synthesized via neutron-capture processes, that can be slow (s) and rapid (r), where slow and rapid is compared to the the $\beta$-decay timescale \citep{Burbidge1957}.} Y, Ba, and La, which are mainly produced by low-mass AGB stars \citep[$\sim$ 1.5--3 M$_\odot$, e.g.,][]{1999ARA&A..37..239B}.
The abundances of Y, Ba, and La were derived as in \citet{2012MNRAS.423.2789D}.
We also gathered Cu abundances, whose 70-- 80\% abundance in the Solar System is ascribed to massive stars
     (\citealt{kappeler11}). We adopted the hyperfine splitting from \citet{1985A&AS...59..403S} and the 
isotopic mixture from \citet{2003AJ....125.2018S} (69\% for $^{63}$Cu and 31\% 
for $^{65}$Cu. We found [Y/Fe] = $-0.15 \pm 0.10$, [Ba/Fe] = $-0.10 \pm 0.10$,
[La/Fe] = $-0.12 \pm 0.10$, and [Cu/Fe] = $0.00 \pm 0.07$. Errors are those derived 
from the spectral synthesis.

We also measured the abundances of Na and Al because these elements are expected 
to be overabundant in the ejecta of {intermediate-mass} AGB stars \citep{Ventura01}, which is
another potential precursor of HD\,8049\,B. We found [Na/Fe] = $0.05\pm0.04$
and [Al/Fe] = $0.13\pm0.03$, indicating no significant pollution from this kind 
of star.

\section{System properties}
\label{sec:system}
\subsection{Orbital properties}
\label{sub:orbit}

\begin{figure*}[h]
\begin{center}
\includegraphics[height=9cm]{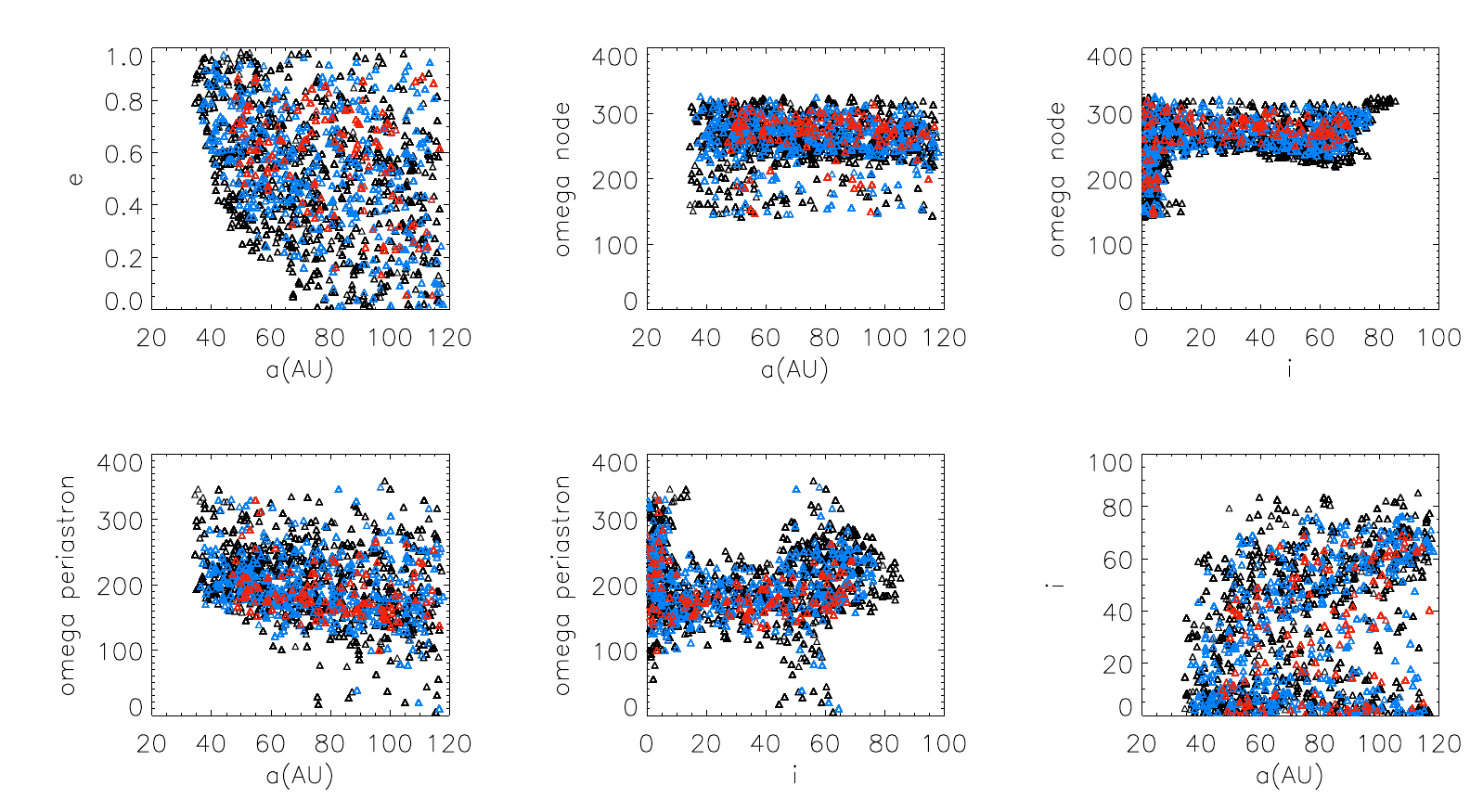}
\caption{{Monte Carlo simulation results showing orbital parameters for orbital solutions compatible with the observational data. Colors indicate different levels of $\chi$-squared values (decreasing from black to blue and finally red). The plots show that some parameters are well constrained in a region of parameters space while others can span all possible values. See \ref{sub:orbit} for information on the Monte Carlo procedures.}}
\label{plotmio}
\end{center}
\end{figure*}

To constrain the possible orbit of the object we performed a Monte
Carlo simulation following the example of \citet{Des11}. The
simulation steps through $1.8\times 10^6$ random orbital elements and rejects all
the orbits that do not fit both the RV trend and the astrometric
data. We adopted $M_{A} = 0.804 \pm 0.023~M_{\odot}$ and $M_{B} = 0.56 \pm 0.08~M_{\odot}$ as described, in Sec.~ \ref{sub:mass} and \ref{sec:wd}. The simulation found $\sim$28000 orbits consistent with the observational data.  We {present the allowed} orbital elements in Figure
\ref{plotmio}.

\begin{figure*}[h]
\begin{center}
\includegraphics[height=11cm]{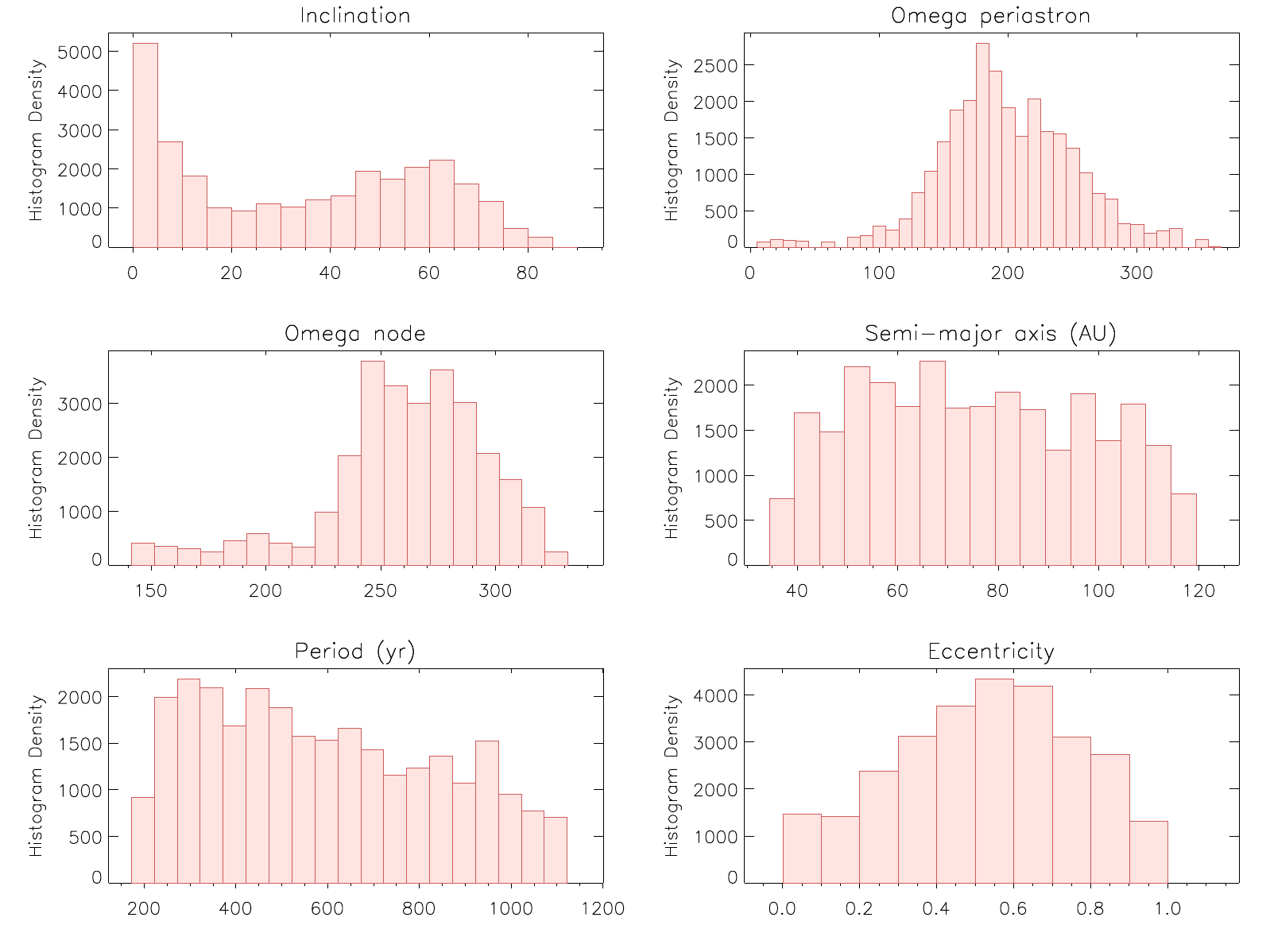}
\caption{Histograms of orbital parameters derived with a Monte Carlo simulation following the example of \citet{Des11}.}
\label{histo}
\end{center}
\end{figure*}

{The} distribution of each element is shown in {the} histograms of
Figure \ref{histo}.  The distribution of the inclinations is bimodal
and has its peaks on the values of $5\deg$ and $65\deg$.  The
argument of periastron $\omega$ and the longitude of the node $\Omega$
have the highest probability around values of $180\deg$ and $250\deg$
respectively. The most probable period is around 300~yr, corresponding
to a semi-major axis of $\sim$ 50~AU. Among the possible configurations of the orbit, there is a lack of edge-on
inclinations that can produce the 
correlation plots of $\omega$ and $\Omega$. This is consistent with the RV trend. The orbit is likely eccentric, with a most probable value $e = 0.55$.

\subsection{Broad band photometry}
\label{sub:photostar}

\begin{table}[h]
\caption{Broad band photometry of HD\,8049} 
\label{t:photom}
\begin{center}       
\begin{tabular}{l l l} 
\hline\hline
Band             &  Mag             & Ref.    \\
\hline
\multicolumn{3}{c}{A+B}\\
\hline
$FUV$              &  $12.645\pm0.004$          &  1      \\
$NUV$              &  $13.454 \pm0.003$          &  1              \\
$FUV-NUV$          &  $-0.809\pm0.005$           & 1       \\
$U$		& $10.098\pm0.031$		&2 \\
$B$		& $9.626\pm0.028$		&2\\
$V$		& $8.734\pm0.027$		& 2 \\
$B-V$              & $+0.892\pm0.007$       &  2  \\
$V-I$              & $+0.95\pm0.02 $       &  3  \\
$U-B$              & $+0.472\pm0.013$             &  2          \\
$J$                & $7.077\pm0.027$  &  4      \\
$H$                & $6.649\pm0.059$  &  4      \\
$K$                & $6.523\pm0.031$  &  4      \\ 
$V-H$              &   $+2.04 \pm0.06$           &   2,4         \\
$V-K$              &   $+2.17 \pm0.04$          &    2,4        \\ 
\hline
\multicolumn{3}{c}{A}\\
\hline
$FUV$& $19.68\pm0.57$  &5\\
$NUV$&  $15.12\pm0.22$  &5\\
$U$		& $10.30\pm0.03$	& 5 \\
$V$                & $8.747\pm0.030$            &   5  \\
$H$           & $6.65\pm0.07$     & 5 \\
\hline
\multicolumn{3}{c}{B}\\
\hline
$FUV$              &  $12.65\pm0.33$           &  5      \\
$NUV$              &  $13.72\pm0.21$           &  5              \\
$U$		& $12.02\pm0.04$ & 5\\
$V$  		& $13.5\pm0.5$ & 6  \\
$H$		& $13.75\pm0.12$ & 5 \\

\hline
\end{tabular}
\tablebib{1: GALEX, AB magnitude system \citep{2005ApJ...619L...1M}
2: \citet{1983SAAOC...7...36C};
3: \citet{1997A&A...323L..49P};
 4: 2MASS \citep{2006AJ....131.1163S} ;
 5: this paper; 
 6: \citet{1929CiUO...80...59V};          
 }

\end{center}
\end{table} 

~\noindent
Table~\ref{t:photom} lists the available broad band photometry for the system.
A very bright $FUV$ magnitude is listed in the GALEX catalog. The offset 
between the optical and GALEX position of $1.67\arcsec$ along the SE direction 
indicates that the $FUV$ source is better coincident with the WD rather 
than the {K2} star.
Such a bright $FUV$ magnitude is not expected from the primary, even
taking into account the {star's} high level of chromospheric activity \citep[see e.g. Fig. 6 in ][]{Smi10}.  Corrections for
chromospheric contribution of the primary have been calculated using
equations 2 and 4 of \citet{Findeisen2011}; we assumed a 
$B-V$ color for the K star as given in Table 15.7 of \citet{Allen}; we assumed $\log R_{\mathrm{HK}}$ = $-4.25$. 
The final derived errors take into account the dispersion of the points used to 
derive the mentioned relations. 

Based on the composite $U$ magnitude of the unresolved system, we determined an individual  $U$ magnitude following the procedures described in Sec.~\ref{sub:uphoto}. $H$ magnitude was obtained with 
NACO observations (see Sec.~\ref{sub:naco}). 

\subsection{System history}
\label{sub:history}

As mentioned in Sec.~\ref{sub:age}, the discrepant results of various age
indicators for HD\,8049\,A might be qualitatively
explained by the accretion of some amounts of mass and angular momentum
from its WD companion HD\,8049\,B.
Comparable cases have been discovered in the past. 
\citet{1996MNRAS.279..180J} showed that the
accretion of slow wind from the AGB progenitor of the WD can transfer
sufficient angular momentum to speed up the companion even for
orbital separation up to 100~AU.  Cases of pairs formed by a WD and
a fast rotating, barium-rich star have been reported by \citet{Jef1996a}.
Another case is represented by the binary {GJ86}, which is known to host a
giant planet and a WD companion \citep[{GJ86}B;][]{Els01,Mug05}.  
{GJ86} also shows anomalies in
the age indicators that are similar, even if less extreme, than those we
found for HD\,8049; the {GJ86} age indicators yield a younger age from X--ray and H\&K than from
kinematic \citep[see][App. B]{Des07}.  
Another more direct indication of wind accretion on a companion during the
AGB phase is represented by the accretion disk around \object{Mira B} 
\citep{1997ApJ...482L.175K,2007ApJ...662..651I}, which
lies at a projected separation of about 70 AU, which is slightly larger than the
most probable value for HD\,8049. 

Following these works we can speculate that part of the material lost 
through stellar wind by the WD progenitor was accreted by HD\,8049\,A, 
increasing its rotation rate and  mimicking a young age for indicators that are directly or indirectly tied to stellar 
rotation (rotation itself, X--ray and H\&K emission). {Specifically,} we expect 
that the activity and rotation age of the system is related to the time 
elapsed since the accretion event, i.e. the cooling age of the WD. However, 
the activity-rotation age might be larger than the cooling age of the WD 
if the primary was not spun-up to the fast rotational velocity typical of 
a very young star. In this scenario, increasing the mass of the {WD} progenitor 
 allows for larger amounts of accreted material, both because of 
large amounts of mass loss due to stellar wind at the end of AGB phase, and also because
of the larger widening of the binary orbit due to mass loss. A more massive progenitor implies also a younger age for the system. On the other hand, 
the lack of significant abundance anomalies on HD\,8049\,A (Sect. 
~\ref{sub:abu}) indicates that either the s-process enhancement of the material lost by the AGB 
progenitor is small, which may arise only for masses lower than 
1.5~M$_{\odot}$, or that the amounts of accreted material is small 
with respect to the heavy element content of the convective zone.

To explore the possible original configuration of the system, we used
the BSE package \citep{Hurley2002}, evolving binary systems with
different primary masses and orbital parameters.  We used {their adopted initial--final mass relationship for the WD and} default
parameters to take into account mass loss and wind accretion. These
elements have significant uncertainty so the results are intended {only} for
a qualitative view of the possible configuration.

A system with a 1.2--1.5~M$_{\odot}$ star initially at 30~AU will produce a
WD of roughly the correct mass with a semi-major axis close to the most probable
value,  for current orbits presented in Sect.~\ref{sub:orbit}. 
The resulting age of the system (3-6 Gyr) 
is compatible with the kinematic properties of HD\,8049. About 0.005~M$_{\odot}$ 
are accreted by the other component in this case, which is about 5\% of 
the mass of the stellar convective zone as derived following the prescriptions 
by \citet{2001ApJ...555..801M}.

{We checked that the results using the more recent semi-empirical relationship
by \citet{Salaris2009} are fully consistent with those derived using BSE.
\citet{Salaris2009} found that WDs with masses in the range 0.56--0.64$~M_{\odot}$ are found in NGC~7789
(estimated age of 1.5~Gyr) and older clusters while those in younger ones, like the Hyades
and Praesepe (age of about 0.65~Gyr), are more massive.} 

\section{Conclusions}
\label{sec:conclusions}

{The substellar candidate around HD\,8049\,A, initially suspected based on NACO $H$-band photometry and astrometry, turned out to be a WD companion rather that a BD.}

With direct imaging, second epoch observations of the object, we confirmed a bound {companionship} and determined {the WD} mass 
through {its} luminosity.

The $M/L$ relation for exoplanets strongly depends on the age of the system 
\citep[see e.g. models from][]{Baraffe2003, Madhusudhan2011}. For this reason,
samples of stars scrutinized for low mass companions through high contrast imaging
usually include a high fraction of presumably young stars. However, a misinterpretation of the age of the host 
star may produce false alarms and/or an erroneous calculation of the mass of the 
companion. This was the case of HD\,8049\,A, a star which appears rejuvenated by 
the accretion of significant angular momentum from the ejecta of the progenitor 
of a WD. The faint companion was confused with a 35~\MJup brown dwarf when the age 
indicated by rotation and activity was considered.

From radial velocity analysis we found a trend that was compatible with a more 
massive object rather than a brown dwarf. {Indeed, a comprehensive analysis of age indicators revealed inconsistencies.} The final clue 
that unequivocally confirms {the WD} hypothesis {is} the observation of the system in $U$- 
band. At this wavelength, a bright object is observed at the same astrometric position of the NACO 
candidate. The ultraviolet magnitude of this object is 1.7~mag fainter than the 
host star.

We found parameters for the WD assuming an atmosphere made of pure hydrogen and fitting photometry observation points. Results are: $T_{\mathrm{eff}} = 18800 \pm 2100$~K, $M_{\mathrm{WD}} = 0.56 \pm 0.08~M_{\odot}$, $\log g = 7.86 \pm 0.16$ and age $t_{\mathrm{cool}} = 63^{+23}_{-17}$ Myr.

We made a Monte Carlo simulation, following the example of \citet{Des11}, to 
constrain the orbital parameters. We found that the WD moves in {a likely} eccentric 
orbit with a most probable period of 300~yr ($\sim$ 50~AU).

In our scenario, the existence of the WD companion (with $H$-band magnitude
compatible with a sub-stellar object) and the spurious young age of the star, as 
derived from rotational period,  $\log R_{\mathrm{HK}}$ and X-ray 
emission, are intrinsically linked.
As discussed in this paper, a way to identify the spurious young stars that have been
 affected by wind accretion is to compare the ages resulting from several 
age indicators, paying attention to discrepancies between rotation-activity ages and lithium 
and kinematic ages. In {the} case {that} the sample is focused on young stars, {such} a 
false alarm is related to a WD with a short cooling age, {which} explain{s} a young
rotation-activity age for {the} companion. Such objects are most easily identified 
in the UV bandpass, as done in our study exploiting archive GALEX observations.

Resolved systems with {white dwarf} companions close enough to have harboured some 
accretion phenomena are interesting targets {for} further constrain{ing} the wind 
accretion occurring in moderately wide binaries during the asymptotic giant
branch (AGB) phase of the WD progenitor.  They also enable investigations of  the maximum binary 
separation at which Ba-stars can be observed. Finally, our results confirm that 
there is a population of WD companions awaiting discovery even at small distance 
from the Sun, as suspected by \citet{2009JPhCS.172a2022H}, and that AO imaging 
can play a relevant role for unveiling it.

As nearby young stars are prime targets for direct imaging planet searches,
and new instrumentation optimized to this science goal is about to start 
operations, one might ask what is the expected frequency of the kind of false 
alarm we {have} found in NACO-LP.  To this purpose, we may first consider the frequency of Ba-stars, which are systems 
similar to those considered here, but with initially smaller separation and shorter 
period. Ba-stars are the result of mass transfer from an AGB companion (the original 
primary of the system) to the small mass star that is currently observed.  For our considered scenarios, the 
system is observed at a late stage in its development,  when the AGB star has evolved into a WD.
From a survey of $\sim$200 slightly evolved F dwarfs, \citet{Tomkin1989} discovered 
two barium dwarfs, and derived a rough frequency of $\sim$1\% Ba-stars among F 
dwarfs; \citet{Edvardsson1993} found 6 dwarf barium stars among 200 F stars (i.e. 
3\%), but their binary nature remains to be proven. \citet{North1991} found a 
frequency of 0.5 to 1.0\% Ba dwarfs among FV stars. \citet{McConnell1972} 
estimated that 1\% of the G and K giants in the vicinity of the Sun are Ba-stars. 
From a 200 star sample, \citet{Williams1975} derived a frequency of Ba-stars 
among G-K giants between 1 and 4\%. \citet{Bidelman1985} obtained a frequency 
between 0.5 and 1\%. Most Ba-stars are in binary systems with orbital 
periods typically longer than 100~days but less than 10~years \citep{McClure1980, 
McClure1984, McClure1990}, though \citet{Jorissen1998} reported a few cases with periods 
outside this range. Orbits typically have low but non-zero 
eccentricity, linked to the fact that they are likely not produced by Roche lobe 
overflow; the origin is probably due to stellar wind. This is further supported by 
evolutionary models \citep{Han1995}.

We may consider the period and secondary mass distribution of binary stars, 
and estimate the fraction of systems that are of potential interest. Integrating 
the distribution of binary separation by \citet{Du1991}, it comes out that at least 
half of the binaries containing a WD have a separation large enough that the star 
does not become a Ba-star. Hence, the fraction of G-K stars with a WD companion 
should be of the order of a few percent. This is consistent with arguments based 
on the initial mass function: roughly a few percent of the G-K stars should 
have a companion originally more massive if we assume that the mass of the two stars
is independent of each other, which is probably a good approximation at least for
solar type stars. Objects like HD\,8049\,A should therefore most likely be G-K stars with a WD companion 
and an original separation typically slightly larger than that required to become {a} Ba-star. 
They should also have experienced some mass and angular momentum transfer.  Hence there should not be large 
separations. 

Objects like HD\,8049\,A may very well pass through the pre-selection of the direct 
imaging candidates, failing to show up as spectroscopic binaries. Periods are long 
{enough} that RV variations are very small.  Furthermore, they will not show up in astrometric binary catalogs, 
because periods are long {that} orbital motion is not distinguished from proper 
motion.  Finally, they might also be overlooked as visual binaries because the contrast is 
large. Inclusion {in} the sample of direct-imaging planet searches also depend{s} on the adopted age
limits. For surveys focused on young stars, such as our NACO {one}, only companions of WD{s} 
with short cooling time{s} might be selected. 

\begin{acknowledgements}
We are grateful to all people involved during the observations at VLT, NTT, 3.6m 
telescope {and 1.2m Euler Swiss telescope at La
Silla}. We also thank the TASTE group of Padua for the precious help 
for EFOSC photometry and Brian Lemaux for his precious help. A.Z., S.D, and  D.M. ackowledge partial support from PRIN INAF 
2010 ``Planetary systems at young ages''. J.C. ackowledges the support of U.S. National 
Science Foundation under Award No. 1009203. Balmer/Lyman lines in the models were 
calculated with the modified Stark broadening profiles of \citet{ber2}, kindly made available by the authors. We acknowledge support from the 
French National Research Agency (ANR) through the GUEPARD project grant ANR10-BLANC0504-01.

We wish to thank the referee for the constructive comments to the paper.
 
\end{acknowledgements}

\bibliographystyle{aa}
\bibliography{hd8049}

\begin{thebibliography}{106}
\expandafter\ifx\csname natexlab\endcsname\relax\def\natexlab#1{#1}\fi

\bibitem[{{Abuter} {et~al.}(2006){Abuter}, {Schreiber}, {Eisenhauer}, {Ott},
  {Horrobin}, \& {Gillesen}}]{Abuter2006}
{Abuter}, R., {Schreiber}, J., {Eisenhauer}, F., {et~al.} 2006, \nar, 50, 398

\bibitem[{{Allen}(1977)}]{Allen}
{Allen}, K.~W. 1977, {Astrophysical quantities.}

\bibitem[{{Baraffe} {et~al.}(2003){Baraffe}, {Chabrier}, {Barman}, {Allard}, \&
  {Hauschildt}}]{Baraffe2003}
{Baraffe}, I., {Chabrier}, G., {Barman}, T.~S., {Allard}, F., \& {Hauschildt},
  P.~H. 2003, \aap, 402, 701

\bibitem[{{Baranne} {et~al.}(1996){Baranne}, {Queloz}, {Mayor}, {Adrianzyk},
  {Knispel}, {Kohler}, {Lacroix}, {Meunier}, {Rimbaud}, \& {Vin}}]{Baranne1996}
{Baranne}, A., {Queloz}, D., {Mayor}, M., {et~al.} 1996, \aaps, 119, 373

\bibitem[{{Barros} {et~al.}(2013){Barros}, {Boue}, {Gibson}, {Pollacco},
  {Santerne}, {Keenan}, {Skillen}, \& {Street}}]{Barros2013}
{Barros}, S.~C.~C., {Boue}, G., {Gibson}, N.~P., {et~al.} 2013, ArXiv e-prints

\bibitem[{{Bayo} {et~al.}(2008){Bayo}, {Rodrigo}, {Barrado Y Navascu{\'e}s},
  {Solano}, {Guti{\'e}rrez}, {Morales-Calder{\'o}n}, \& {Allard}}]{Bayo2008}
{Bayo}, A., {Rodrigo}, C., {Barrado Y Navascu{\'e}s}, D., {et~al.} 2008, \aap,
  492, 277

\bibitem[{{Bergeron} {et~al.}(2011){Bergeron}, {Wesemael}, {Dufour},
  {Beauchamp}, {Hunter}, {Saffer}, {Gianninas}, {Ruiz}, {Limoges}, {Dufour},
  {Fontaine}, \& {Liebert}}]{2011ApJ...737...28B}
{Bergeron}, P., {Wesemael}, F., {Dufour}, P., {et~al.} 2011, \apj, 737, 28

\bibitem[{{Biazzo} {et~al.}(2012){Biazzo}, {D'Orazi}, {Desidera}, {Covino},
  {Alcal{\'a}}, \& {Zusi}}]{Biazzo2012}
{Biazzo}, K., {D'Orazi}, V., {Desidera}, S., {et~al.} 2012, \mnras, 427, 2905

\bibitem[{{Bidelman}(1985)}]{Bidelman1985}
{Bidelman}, W.~P. 1985, in Astrophysics and Space Science Library, Vol. 114,
  Cool Stars with Excesses of Heavy Elements, ed. M.~{Jaschek} \& P.~C.
  {Keenan}, 43--45

\bibitem[{{Boisse} {et~al.}(2011){Boisse}, {Bouchy}, {H{\'e}brard}, {Bonfils},
  {Santos}, \& {Vauclair}}]{2011A&A...528A...4B}
{Boisse}, I., {Bouchy}, F., {H{\'e}brard}, G., {et~al.} 2011, \aap, 528, A4

\bibitem[{{Bonnet} {et~al.}(2004{\natexlab{a}}){Bonnet}, {Abuter}, {Baker},
  {Bornemann}, {Brown}, {Castillo}, {Conzelmann}, {Damster}, {Davies},
  {Delabre}, {Donaldson}, {Dumas}, {Eisenhauer}, {Elswijk}, {Fedrigo},
  {Finger}, {Gemperlein}, {Genzel}, {Gilbert}, {Gillet}, {Goldbrunner},
  {Horrobin}, {Ter Horst}, {Huber}, {Hubin}, {Iserlohe}, {Kaufer},
  {Kissler-Patig}, {Kragt}, {Kroes}, {Lehnert}, {Lieb}, {Liske}, {Lizon},
  {Lutz}, {Modigliani}, {Monnet}, {Nesvadba}, {Patig}, {Pragt}, {Reunanen},
  {R{\"o}hrle}, {Rossi}, {Schmutzer}, {Schoenmaker}, {Schreiber}, {Stroebele},
  {Szeifert}, {Tacconi}, {Tecza}, {Thatte}, {Tordo}, {van der Werf}, \&
  {Weisz}}]{Bonnet2004b}
{Bonnet}, H., {Abuter}, R., {Baker}, A., {et~al.} 2004{\natexlab{a}}, The
  Messenger, 117, 17

\bibitem[{{Bonnet} {et~al.}(2004{\natexlab{b}}){Bonnet}, {Conzelmann},
  {Delabre}, {Donaldson}, {Fedrigo}, {Hubin}, {Kissler-Patig}, {Lizon},
  {Paufique}, {Rossi}, {Stroebele}, \& {Tordo}}]{Bonnet2004a}
{Bonnet}, H., {Conzelmann}, R., {Delabre}, B., {et~al.} 2004{\natexlab{b}}, in
  Society of Photo-Optical Instrumentation Engineers (SPIE) Conference Series,
  Vol. 5490, Society of Photo-Optical Instrumentation Engineers (SPIE)
  Conference Series, ed. D.~{Bonaccini Calia}, B.~L. {Ellerbroek}, \&
  R.~{Ragazzoni}, 130--138

\bibitem[{{Bonnet} {et~al.}(2003){Bonnet}, {Str{\"o}bele}, {Biancat-Marchet},
  {Brynnel}, {Conzelmann}, {Delabre}, {Donaldson}, {Farinato}, {Fedrigo},
  {Hubin}, {Kasper}, \& {Kissler-Patig}}]{Bonnet2003}
{Bonnet}, H., {Str{\"o}bele}, S., {Biancat-Marchet}, F., {et~al.} 2003, in
  Society of Photo-Optical Instrumentation Engineers (SPIE) Conference Series,
  Vol. 4839, Society of Photo-Optical Instrumentation Engineers (SPIE)
  Conference Series, ed. P.~L. {Wizinowich} \& D.~{Bonaccini}, 329--343

\bibitem[{{Brown}(2003)}]{Brown2003}
{Brown}, T.~M. 2003, \apjl, 593, L125

\bibitem[{{Burbidge} {et~al.}(1957){Burbidge}, {Burbidge}, {Fowler}, \&
  {Hoyle}}]{Burbidge1957}
{Burbidge}, E.~M., {Burbidge}, G.~R., {Fowler}, W.~A., \& {Hoyle}, F. 1957,
  Reviews of Modern Physics, 29, 547

\bibitem[{{Busso} {et~al.}(1999){Busso}, {Gallino}, \&
  {Wasserburg}}]{1999ARA&A..37..239B}
{Busso}, M., {Gallino}, R., \& {Wasserburg}, G.~J. 1999, \araa, 37, 239

\bibitem[{{Butters} {et~al.}(2010){Butters}, {West}, {Anderson}, {Collier
  Cameron}, {Clarkson}, {Enoch}, {Haswell}, {Hellier}, {Horne}, {Joshi},
  {Kane}, {Lister}, {Maxted}, {Parley}, {Pollacco}, {Smalley}, {Street},
  {Todd}, {Wheatley}, \& {Wilson}}]{Butters2010}
{Butters}, O.~W., {West}, R.~G., {Anderson}, D.~R., {et~al.} 2010, \aap, 520,
  L10

\bibitem[{{Chauvin}(2010)}]{chauvin2010}
{Chauvin}, G. 2010, in In the Spirit of Lyot 2010

\bibitem[{{Chauvin} {et~al.}(2010){Chauvin}, {Lagrange}, {Bonavita},
  {Zuckerman}, {Dumas}, {Bessell}, {Beuzit}, {Bonnefoy}, {Desidera}, {Farihi},
  {Lowrance}, {Mouillet}, \& {Song}}]{Chauvin2}
{Chauvin}, G., {Lagrange}, A.-M., {Bonavita}, M., {et~al.} 2010, \aap, 509, A52

\bibitem[{{Chauvin} {et~al.}(2007){Chauvin}, {Lagrange}, {Udry}, \&
  {Mayor}}]{Chauvin2007}
{Chauvin}, G., {Lagrange}, A.-M., {Udry}, S., \& {Mayor}, M. 2007, \aap, 475,
  723

\bibitem[{{Cousins}(1983)}]{1983SAAOC...7...36C}
{Cousins}, A.~W.~J. 1983, South African Astronomical Observatory Circular, 7,
  36

\bibitem[{{da Silva} {et~al.}(2006){da Silva}, {Girardi}, {Pasquini},
  {Setiawan}, {von der L{\"u}he}, {de Medeiros}, {Hatzes}, {D{\"o}llinger}, \&
  {Weiss}}]{2006A&A...458..609D}
{da Silva}, L., {Girardi}, L., {Pasquini}, L., {et~al.} 2006, \aap, 458, 609

\bibitem[{{Delorme} {et~al.}(2011){Delorme}, {Collier Cameron}, {Hebb},
  {Rostron}, {Lister}, {Norton}, {Pollacco}, \& {West}}]{Delorme2011}
{Delorme}, P., {Collier Cameron}, A., {Hebb}, L., {et~al.} 2011, \mnras, 413,
  2218

\bibitem[{{Desidera} \& {Barbieri}(2007)}]{Des07}
{Desidera}, S. \& {Barbieri}, M. 2007, \aap, 462, 345

\bibitem[{{Desidera} {et~al.}(2011){Desidera}, {Carolo}, {Gratton}, {Martinez
  Fiorenzano}, {Endl}, {Mesa}, {Barbieri}, {Bonavita}, {Cecconi}, {Claudi},
  {Cosentino}, {Marzari}, \& {Scuderi}}]{Des11}
{Desidera}, S., {Carolo}, E., {Gratton}, R., {et~al.} 2011, \aap, 533, A90

\bibitem[{{Dobbie} {et~al.}(2005){Dobbie}, {Burleigh}, {Levan}, {Barstow},
  {Napiwotzki}, {Holberg}, {Hubeny}, \& {Howell}}]{Dobbie2005}
{Dobbie}, P.~D., {Burleigh}, M.~R., {Levan}, A.~J., {et~al.} 2005, \mnras, 357,
  1049

\bibitem[{{D'Orazi} {et~al.}(2012){D'Orazi}, {Biazzo}, {Desidera}, {Covino},
  {Andrievsky}, \& {Gratton}}]{2012MNRAS.423.2789D}
{D'Orazi}, V., {Biazzo}, K., {Desidera}, S., {et~al.} 2012, \mnras, 423, 2789

\bibitem[{{Dumusque} {et~al.}(2012){Dumusque}, {Pepe}, {Lovis},
  {S{\'e}gransan}, {Sahlmann}, {Benz}, {Bouchy}, {Mayor}, {Queloz}, {Santos},
  \& {Udry}}]{2012Natur.491..207D}
{Dumusque}, X., {Pepe}, F., {Lovis}, C., {et~al.} 2012, \nat, 491, 207

\bibitem[{{Duquennoy} \& {Mayor}(1991)}]{Du1991}
{Duquennoy}, A. \& {Mayor}, M. 1991, \aap, 248, 485

\bibitem[{{Edvardsson} {et~al.}(1993){Edvardsson}, {Andersen}, {Gustafsson},
  {Lambert}, {Nissen}, \& {Tomkin}}]{Edvardsson1993}
{Edvardsson}, B., {Andersen}, J., {Gustafsson}, B., {et~al.} 1993, \aap, 275,
  101

\bibitem[{{Eisenhauer} {et~al.}(2003){Eisenhauer}, {Abuter}, {Bickert},
  {Biancat-Marchet}, {Bonnet}, {Brynnel}, {Conzelmann}, {Delabre}, {Donaldson},
  {Farinato}, {Fedrigo}, {Genzel}, {Hubin}, {Iserlohe}, {Kasper},
  {Kissler-Patig}, {Monnet}, {Roehrle}, {Schreiber}, {Stroebele}, {Tecza},
  {Thatte}, \& {Weisz}}]{Eisenhauer2003}
{Eisenhauer}, F., {Abuter}, R., {Bickert}, K., {et~al.} 2003, in Society of
  Photo-Optical Instrumentation Engineers (SPIE) Conference Series, Vol. 4841,
  Society of Photo-Optical Instrumentation Engineers (SPIE) Conference Series,
  ed. M.~{Iye} \& A.~F.~M. {Moorwood}, 1548--1561

\bibitem[{{Els} {et~al.}(2001){Els}, {Sterzik}, {Marchis}, {Pantin}, {Endl}, \&
  {K{\"u}rster}}]{Els01}
{Els}, S.~G., {Sterzik}, M.~F., {Marchis}, F., {et~al.} 2001, \aap, 370, L1

\bibitem[{{Figueira} {et~al.}(2010){Figueira}, {Marmier}, {Bonfils}, {di
  Folco}, {Udry}, {Santos}, {Lovis}, {M{\'e}gevand}, {Melo}, {Pepe}, {Queloz},
  {S{\'e}gransan}, {Triaud}, \& {Viana Almeida}}]{2010A&A...513L...8F}
{Figueira}, P., {Marmier}, M., {Bonfils}, X., {et~al.} 2010, \aap, 513, L8

\bibitem[{{Findeisen} {et~al.}(2011){Findeisen}, {Hillenbrand}, \&
  {Soderblom}}]{Findeisen2011}
{Findeisen}, K., {Hillenbrand}, L., \& {Soderblom}, D. 2011, \aj, 142, 23

\bibitem[{{Fitzpatrick}(1999)}]{Fiz1999}
{Fitzpatrick}, E.~L. 1999, \pasp, 111, 63

\bibitem[{{Giammichele} {et~al.}(2012){Giammichele}, {Bergeron}, \&
  {Dufour}}]{Giammichele2012}
{Giammichele}, N., {Bergeron}, P., \& {Dufour}, P. 2012, \apjs, 199, 29

\bibitem[{{Gould} {et~al.}(2012){Gould}, {Yee}, {Bond}, {Udalski}, {Han},
  {Jorgensen}, {Greenhill}, {Tsapras}, {Pinsonneault}, {Bensby}, {uFUN}, {MOA},
  {OGLE}, {MiNDSTEp}, {RoboNet}, \& {PLANET collaborations}}]{Gould2012}
{Gould}, A., {Yee}, J.~C., {Bond}, I.~A., {et~al.} 2012, ArXiv e-prints

\bibitem[{{Gray} {et~al.}(2006){Gray}, {Corbally}, {Garrison}, {McFadden},
  {Bubar}, {McGahee}, {O'Donoghue}, \& {Knox}}]{Gray2006}
{Gray}, R.~O., {Corbally}, C.~J., {Garrison}, R.~F., {et~al.} 2006, \aj, 132,
  161

\bibitem[{{Hamuy} {et~al.}(1994){Hamuy}, {Suntzeff}, {Heathcote}, {Walker},
  {Gigoux}, \& {Phillips}}]{Hamuy1994}
{Hamuy}, M., {Suntzeff}, N.~B., {Heathcote}, S.~R., {et~al.} 1994, \pasp, 106,
  566

\bibitem[{{Han} {et~al.}(1995){Han}, {Eggleton}, {Podsiadlowski}, \&
  {Tout}}]{Han1995}
{Han}, Z., {Eggleton}, P.~P., {Podsiadlowski}, P., \& {Tout}, C.~A. 1995,
  \mnras, 277, 1443

\bibitem[{{Heintz}(2000)}]{Heintz}
{Heintz}, W. 2000, {Visual Binary Stars}, ed. P.~{Murdin}

\bibitem[{{Hern{\'a}n-Obispo} {et~al.}(2010){Hern{\'a}n-Obispo},
  {G{\'a}lvez-Ortiz}, {Anglada-Escud{\'e}}, {Kane}, {Barnes}, {de Castro}, \&
  {Cornide}}]{2010A&A...512A..45H}
{Hern{\'a}n-Obispo}, M., {G{\'a}lvez-Ortiz}, M.~C., {Anglada-Escud{\'e}}, G.,
  {et~al.} 2010, \aap, 512, A45

\bibitem[{{Holberg}(2009)}]{2009JPhCS.172a2022H}
{Holberg}, J.~B. 2009, Journal of Physics Conference Series, 172, 012022

\bibitem[{{Holberg} \& {Bergeron}(2006)}]{2006AJ....132.1221H}
{Holberg}, J.~B. \& {Bergeron}, P. 2006, \aj, 132, 1221

\bibitem[{{Holmberg} {et~al.}(2009){Holmberg}, {Nordstr{\"o}m}, \&
  {Andersen}}]{Holmberg}
{Holmberg}, J., {Nordstr{\"o}m}, B., \& {Andersen}, J. 2009, \aap, 501, 941

\bibitem[{{Hu{\'e}lamo} {et~al.}(2008){Hu{\'e}lamo}, {Figueira}, {Bonfils},
  {Santos}, {Pepe}, {Gillon}, {Azevedo}, {Barman}, {Fern{\'a}ndez}, {di Folco},
  {Guenther}, {Lovis}, {Melo}, {Queloz}, \& {Udry}}]{2008A&A...489L...9H}
{Hu{\'e}lamo}, N., {Figueira}, P., {Bonfils}, X., {et~al.} 2008, \aap, 489, L9

\bibitem[{{Hurley} {et~al.}(2002){Hurley}, {Tout}, \& {Pols}}]{Hurley2002}
{Hurley}, J.~R., {Tout}, C.~A., \& {Pols}, O.~R. 2002, \mnras, 329, 897

\bibitem[{{Ireland} {et~al.}(2007){Ireland}, {Monnier}, {Tuthill}, {Cohen}, {De
  Buizer}, {Packham}, {Ciardi}, {Hayward}, \& {Lloyd}}]{2007ApJ...662..651I}
{Ireland}, M.~J., {Monnier}, J.~D., {Tuthill}, P.~G., {et~al.} 2007, \apj, 662,
  651

\bibitem[{{Jeffries} \& {Smalley}(1996)}]{Jef1996a}
{Jeffries}, R.~D. \& {Smalley}, B. 1996, \aap, 315, L19

\bibitem[{{Jeffries} \& {Stevens}(1996)}]{1996MNRAS.279..180J}
{Jeffries}, R.~D. \& {Stevens}, I.~R. 1996, \mnras, 279, 180

\bibitem[{{Jorissen} {et~al.}(1998){Jorissen}, {Van Eck}, {Mayor}, \&
  {Udry}}]{Jorissen1998}
{Jorissen}, A., {Van Eck}, S., {Mayor}, M., \& {Udry}, S. 1998, \aap, 332, 877

\bibitem[{{K{\"a}ppeler} {et~al.}(2011){K{\"a}ppeler}, {Gallino}, {Bisterzo},
  \& {Aoki}}]{kappeler11}
{K{\"a}ppeler}, F., {Gallino}, R., {Bisterzo}, S., \& {Aoki}, W. 2011, Reviews
  of Modern Physics, 83, 157

\bibitem[{{Karovska} {et~al.}(1997){Karovska}, {Hack}, {Raymond}, \&
  {Guinan}}]{1997ApJ...482L.175K}
{Karovska}, M., {Hack}, W., {Raymond}, J., \& {Guinan}, E. 1997, \apjl, 482,
  L175

\bibitem[{{Koester}(2010)}]{Koester2010}
{Koester}, D. 2010, \memsai, 81, 921

\bibitem[{{Kowalski} \& {Saumon}(2006)}]{2006ApJ...651L.137K}
{Kowalski}, P.~M. \& {Saumon}, D. 2006, \apjl, 651, L137

\bibitem[{{Lafreni{\`e}re} {et~al.}(2007){Lafreni{\`e}re}, {Marois}, {Doyon},
  {Nadeau}, \& {Artigau}}]{lafreniere2007}
{Lafreni{\`e}re}, D., {Marois}, C., {Doyon}, R., {Nadeau}, D., \& {Artigau},
  {\'E}. 2007, \apj, 660, 770

\bibitem[{{Lanza} {et~al.}(2010){Lanza}, {Bonomo}, {Moutou}, {Pagano},
  {Messina}, {Leto}, {Cutispoto}, {Aigrain}, {Alonso}, {Barge}, {Deleuil},
  {Auvergne}, {Baglin}, \& {Collier Cameron}}]{2010A&A...520A..53L}
{Lanza}, A.~F., {Bonomo}, A.~S., {Moutou}, C., {et~al.} 2010, \aap, 520, A53

\bibitem[{{Lenzen} {et~al.}(2003){Lenzen}, {Hartung}, {Brandner}, {Finger},
  {Hubin}, {Lacombe}, {Lagrange}, {Lehnert}, {Moorwood}, \&
  {Mouillet}}]{lenzen2003}
{Lenzen}, R., {Hartung}, M., {Brandner}, W., {et~al.} 2003, in {SPIE Conference
  Series}, Vol. 4841, 944--952

\bibitem[{{Lovis} {et~al.}(2011){Lovis}, {Dumusque}, {Santos}, {Bouchy},
  {Mayor}, {Pepe}, {Queloz}, {S{\'e}gransan}, \& {Udry}}]{2011arXiv1107.5325L}
{Lovis}, C., {Dumusque}, X., {Santos}, N.~C., {et~al.} 2011, ArXiv e-prints

\bibitem[{{Madhusudhan} {et~al.}(2011){Madhusudhan}, {Burrows}, \&
  {Currie}}]{Madhusudhan2011}
{Madhusudhan}, N., {Burrows}, A., \& {Currie}, T. 2011, \apj, 737, 34

\bibitem[{{Mamajek} \& {Hillenbrand}(2008)}]{Mam08}
{Mamajek}, E.~E. \& {Hillenbrand}, L.~A. 2008, \apj, 687, 1264

\bibitem[{{Marois} {et~al.}(2006){Marois}, {Lafreni{\`e}re}, {Doyon},
  {Macintosh}, \& {Nadeau}}]{marois2006}
{Marois}, C., {Lafreni{\`e}re}, D., {Doyon}, R., {Macintosh}, B., \& {Nadeau},
  D. 2006, \apj, 641, 556

\bibitem[{{Martin} {et~al.}(2005){Martin}, {Fanson}, {Schiminovich},
  {Morrissey}, {Friedman}, {Barlow}, {Conrow}, {Grange}, {Jelinsky},
  {Milliard}, {Siegmund}, {Bianchi}, {Byun}, {Donas}, {Forster}, {Heckman},
  {Lee}, {Madore}, {Malina}, {Neff}, {Rich}, {Small}, {Surber}, {Szalay},
  {Welsh}, \& {Wyder}}]{2005ApJ...619L...1M}
{Martin}, D.~C., {Fanson}, J., {Schiminovich}, D., {et~al.} 2005, \apjl, 619,
  L1

\bibitem[{{Mason} {et~al.}(2001){Mason}, {Wycoff}, {Hartkopf}, {Douglass}, \&
  {Worley}}]{wds}
{Mason}, B.~D., {Wycoff}, G.~L., {Hartkopf}, W.~I., {Douglass}, G.~G., \&
  {Worley}, C.~E. 2001, \aj, 122, 3466

\bibitem[{{Mawet} {et~al.}(2012){Mawet}, {Absil}, {Montagnier}, {Riaud},
  {Surdej}, {Ducourant}, {Augereau}, {R{\"o}ttinger}, {Girard}, {Krist}, \&
  {Stapelfeldt}}]{Maw12}
{Mawet}, D., {Absil}, O., {Montagnier}, G., {et~al.} 2012, \aap, 544, A131

\bibitem[{{McClure}(1984)}]{McClure1984}
{McClure}, R.~D. 1984, \pasp, 96, 117

\bibitem[{{McClure} {et~al.}(1980){McClure}, {Fletcher}, \&
  {Nemec}}]{McClure1980}
{McClure}, R.~D., {Fletcher}, J.~M., \& {Nemec}, J.~M. 1980, \apjl, 238, L35

\bibitem[{{McClure} \& {Woodsworth}(1990)}]{McClure1990}
{McClure}, R.~D. \& {Woodsworth}, A.~W. 1990, \apj, 352, 709

\bibitem[{{McConnell} \& {Gast}(1972)}]{McConnell1972}
{McConnell}, R.~K. \& {Gast}, P.~W. 1972, Moon, 5, 41

\bibitem[{{McCrea}(1964)}]{McCrea1964}
{McCrea}, W.~H. 1964, \mnras, 128, 335

\bibitem[{{Messina} {et~al.}(2011){Messina}, {Desidera}, {Lanzafame},
  {Turatto}, \& {Guinan}}]{Mes11}
{Messina}, S., {Desidera}, S., {Lanzafame}, A.~C., {Turatto}, M., \& {Guinan},
  E.~F. 2011, \aap, 532, A10

\bibitem[{{Messina} {et~al.}(2010){Messina}, {Desidera}, {Turatto},
  {Lanzafame}, \& {Guinan}}]{Messina2010}
{Messina}, S., {Desidera}, S., {Turatto}, M., {Lanzafame}, A.~C., \& {Guinan},
  E.~F. 2010, \aap, 520, A15

\bibitem[{{Metchev} \& {Hillenbrand}(2009)}]{2009ApJS..181...62M}
{Metchev}, S.~A. \& {Hillenbrand}, L.~A. 2009, \apjs, 181, 62

\bibitem[{{Montes} {et~al.}(2001){Montes}, {L{\'o}pez-Santiago}, {G{\'a}lvez},
  {Fern{\'a}ndez-Figueroa}, {De Castro}, \& {Cornide}}]{Montes2001}
{Montes}, D., {L{\'o}pez-Santiago}, J., {G{\'a}lvez}, M.~C., {et~al.} 2001,
  \mnras, 328, 45

\bibitem[{{Mugrauer} \& {Neuh{\"a}user}(2005)}]{Mug05}
{Mugrauer}, M. \& {Neuh{\"a}user}, R. 2005, \mnras, 361, L15

\bibitem[{{Murray} {et~al.}(2001){Murray}, {Chaboyer}, {Arras}, {Hansen}, \&
  {Noyes}}]{2001ApJ...555..801M}
{Murray}, N., {Chaboyer}, B., {Arras}, P., {Hansen}, B., \& {Noyes}, R.~W.
  2001, \apj, 555, 801

\bibitem[{{North} \& {Duquennoy}(1991)}]{North1991}
{North}, P. \& {Duquennoy}, A. 1991, \aap, 244, 335

\bibitem[{{O'Donovan} {et~al.}(2006){O'Donovan}, {Charbonneau}, {Torres},
  {Mandushev}, {Dunham}, {Latham}, {Alonso}, {Brown}, {Esquerdo}, {Everett}, \&
  {Creevey}}]{Odo06}
{O'Donovan}, F.~T., {Charbonneau}, D., {Torres}, G., {et~al.} 2006, \apj, 644,
  1237

\bibitem[{{Perryman} {et~al.}(1997){Perryman}, {Lindegren}, {Kovalevsky},
  {Hoeg}, {Bastian}, {Bernacca}, {Cr{\'e}z{\'e}}, {Donati}, {Grenon},
  {Grewing}, {van Leeuwen}, {van der Marel}, {Mignard}, {Murray}, {Le Poole},
  {Schrijver}, {Turon}, {Arenou}, {Froeschl{\'e}}, \&
  {Petersen}}]{1997A&A...323L..49P}
{Perryman}, M.~A.~C., {Lindegren}, L., {Kovalevsky}, J., {et~al.} 1997, \aap,
  323, L49

\bibitem[{{Pickles}(1998)}]{Pickles1998}
{Pickles}, A.~J. 1998, VizieR Online Data Catalog, 611, 863

\bibitem[{{Pojmanski}(2002)}]{Pojmaski2002}
{Pojmanski}, G. 2002, \actaa, 52, 397

\bibitem[{{Pollacco} {et~al.}(2006){Pollacco}, {Skillen}, {Collier Cameron},
  {Christian}, {Hellier}, {Irwin}, {Lister}, {Street}, {West}, {Anderson},
  {Clarkson}, {Deeg}, {Enoch}, {Evans}, {Fitzsimmons}, {Haswell}, {Hodgkin},
  {Horne}, {Kane}, {Keenan}, {Maxted}, {Norton}, {Osborne}, {Parley}, {Ryans},
  {Smalley}, {Wheatley}, \& {Wilson}}]{Pollacco2006}
{Pollacco}, D.~L., {Skillen}, I., {Collier Cameron}, A., {et~al.} 2006, \pasp,
  118, 1407

\bibitem[{{Porto de Mello} \& {da Silva}(1997)}]{portodemello1997}
{Porto de Mello}, G.~F. \& {da Silva}, L. 1997, \apjl, 476, L89

\bibitem[{{Queloz} {et~al.}(2001){Queloz}, {Henry}, {Sivan}, {Baliunas},
  {Beuzit}, {Donahue}, {Mayor}, {Naef}, {Perrier}, \& {Udry}}]{Que01}
{Queloz}, D., {Henry}, G.~W., {Sivan}, J.~P., {et~al.} 2001, \aap, 379, 279

\bibitem[{{Rousset} {et~al.}(2003){Rousset}, {Lacombe}, {Puget}, {Hubin},
  {Gendron}, {Fusco}, {Arsenault}, {Charton}, {Feautrier}, {Gigan}, {Kern},
  {Lagrange}, {Madec}, {Mouillet}, {Rabaud}, {Rabou}, {Stadler}, \&
  {Zins}}]{rousset2003}
{Rousset}, G., {Lacombe}, F., {Puget}, P., {et~al.} 2003, in {SPIE Conference
  Series}, ed. P.~L. {Wizinowich} \& D.~{Bonaccini}, Vol. 4839, 140--149

\bibitem[{{Salaris} {et~al.}(2009){Salaris}, {Serenelli}, {Weiss}, \& {Miller
  Bertolami}}]{Salaris2009}
{Salaris}, M., {Serenelli}, A., {Weiss}, A., \& {Miller Bertolami}, M. 2009,
  \apj, 692, 1013

\bibitem[{{Scargle}(1982)}]{Scargle1982}
{Scargle}, J.~D. 1982, \apj, 263, 835

\bibitem[{{Setiawan} {et~al.}(2008){Setiawan}, {Henning}, {Launhardt},
  {M{\"u}ller}, {Weise}, \& {K{\"u}rster}}]{2008Natur.451...38S}
{Setiawan}, J., {Henning}, T., {Launhardt}, R., {et~al.} 2008, \nat, 451, 38

\bibitem[{{Simmerer} {et~al.}(2003){Simmerer}, {Sneden}, {Ivans}, {Kraft},
  {Shetrone}, \& {Smith}}]{2003AJ....125.2018S}
{Simmerer}, J., {Sneden}, C., {Ivans}, I.~I., {et~al.} 2003, \aj, 125, 2018

\bibitem[{{Skrutskie} {et~al.}(2006){Skrutskie}, {Cutri}, {Stiening},
  {Weinberg}, {Schneider}, {Carpenter}, {Beichman}, {Capps}, {Chester},
  {Elias}, {Huchra}, {Liebert}, {Lonsdale}, {Monet}, {Price}, {Seitzer},
  {Jarrett}, {Kirkpatrick}, {Gizis}, {Howard}, {Evans}, {Fowler}, {Fullmer},
  {Hurt}, {Light}, {Kopan}, {Marsh}, {McCallon}, {Tam}, {Van Dyk}, \&
  {Wheelock}}]{2006AJ....131.1163S}
{Skrutskie}, M.~F., {Cutri}, R.~M., {Stiening}, R., {et~al.} 2006, \aj, 131,
  1163

\bibitem[{{Smith} \& {Redenbaugh}(2010)}]{Smi10}
{Smith}, G.~H. \& {Redenbaugh}, A.~K. 2010, \pasp, 122, 1303

\bibitem[{{Steffen}(1985)}]{1985A&AS...59..403S}
{Steffen}, M. 1985, \aaps, 59, 403

\bibitem[{{Tomkin} {et~al.}(1989){Tomkin}, {Lambert}, {Edvardsson},
  {Gustafsson}, \& {Nissen}}]{Tomkin1989}
{Tomkin}, J., {Lambert}, D.~L., {Edvardsson}, B., {Gustafsson}, B., \&
  {Nissen}, P.~E. 1989, \aap, 219, L15

\bibitem[{{Torres} {et~al.}(2006){Torres}, {Quast}, {da Silva}, {de La Reza},
  {Melo}, \& {Sterzik}}]{Tor06}
{Torres}, C.~A.~O., {Quast}, G.~R., {da Silva}, L., {et~al.} 2006, \aap, 460,
  695

\bibitem[{{Torres} {et~al.}(2011){Torres}, {Fressin}, {Batalha}, {Borucki},
  {Brown}, {Bryson}, {Buchhave}, {Charbonneau}, {Ciardi}, {Dunham}, {Fabrycky},
  {Ford}, {Gautier}, {Gilliland}, {Holman}, {Howell}, {Isaacson}, {Jenkins},
  {Koch}, {Latham}, {Lissauer}, {Marcy}, {Monet}, {Prsa}, {Quinn}, {Ragozzine},
  {Rowe}, {Sasselov}, {Steffen}, \& {Welsh}}]{torres2011}
{Torres}, G., {Fressin}, F., {Batalha}, N.~M., {et~al.} 2011, \apj, 727, 24

\bibitem[{Tremblay \& Bergeron(2009)}]{ber2}
Tremblay, P.-E. \& Bergeron, P. 2009, The Astrophysical Journal, 696, 1755

\bibitem[{{Tremblay} {et~al.}(2011){Tremblay}, {Bergeron}, \&
  {Gianninas}}]{2011ApJ...730..128T}
{Tremblay}, P.-E., {Bergeron}, P., \& {Gianninas}, A. 2011, \apj, 730, 128

\bibitem[{{van den Bos}(1929)}]{1929CiUO...80...59V}
{van den Bos}, W.~H. 1929, Circular of the Union Observatory Johannesburg, 80,
  59

\bibitem[{{van Leeuwen}(2007)}]{2007A&A...474..653V}
{van Leeuwen}, F. 2007, \aap, 474, 653

\bibitem[{{Vennes}(1999)}]{1999ApJ...525..995V}
{Vennes}, S. 1999, \apj, 525, 995

\bibitem[{{Vennes} {et~al.}(2011){Vennes}, {Kawka}, \& {N{\'e}meth}}]{Ven11}
{Vennes}, S., {Kawka}, A., \& {N{\'e}meth}, P. 2011, \mnras, 410, 2095

\bibitem[{{Ventura} {et~al.}(2001){Ventura}, {D'Antona}, {Mazzitelli}, \&
  {Gratton}}]{Ventura01}
{Ventura}, P., {D'Antona}, F., {Mazzitelli}, I., \& {Gratton}, R. 2001, \apjl,
  550, L65

\bibitem[{{Vigan} {et~al.}(2012){Vigan}, {Patience}, {Marois}, {Bonavita}, {De
  Rosa}, {Macintosh}, {Song}, {Doyon}, {Zuckerman}, {Lafreni{\`e}re}, \&
  {Barman}}]{vigan2012}
{Vigan}, A., {Patience}, J., {Marois}, C., {et~al.} 2012, \aap, 544, A9

\bibitem[{{Voges} {et~al.}(1999){Voges}, {Aschenbach}, {Boller},
  {Br{\"a}uninger}, {Briel}, {Burkert}, {Dennerl}, {Englhauser}, {Gruber},
  {Haberl}, {Hartner}, {Hasinger}, {K{\"u}rster}, {Pfeffermann}, {Pietsch},
  {Predehl}, {Rosso}, {Schmitt}, {Tr{\"u}mper}, \&
  {Zimmermann}}]{1999A&A...349..389V}
{Voges}, W., {Aschenbach}, B., {Boller}, T., {et~al.} 1999, \aap, 349, 389

\bibitem[{{Williams}(1975)}]{Williams1975}
{Williams}, P.~M. 1975, \mnras, 170, 343

\bibitem[{{Zorotovic} \& {Schreiber}(2012)}]{2012arXiv1211.5356Z}
{Zorotovic}, M. \& {Schreiber}, M.~R. 2012, ArXiv e-prints

\end{thebibliography}

\end{document}